\documentclass[12pt]{article}
\usepackage{graphicx}
\usepackage{subfigure}
\usepackage{algorithm}
\usepackage{algpseudocode}
\usepackage{appendix}
\usepackage{amsmath}
\usepackage{amsfonts}
\usepackage{amsthm}
\usepackage{cancel}
\usepackage{setspace}
\usepackage{url}
\usepackage{authblk}
\usepackage{verbatim,textcomp}
\usepackage{color}

\graphicspath{{./figs/}}

\newtheorem{thm}{Theorem}[section]

\newtheorem{corollary}{Corollary}

\newtheorem{prop}[thm]{Proposition}

\newtheorem{lemma}{Lemma}

\theoremstyle{definition}

\theoremstyle{remark}

\newcommand{\orderof}[1]{{\ensuremath{ {\cal O}\left(#1\right)}}}

\newcommand{\arrL}{\stackrel{{\cal L}}{\longrightarrow}}
\newcommand{\arrP}{\stackrel{{P}}{\longrightarrow}}

\newcommand{\finvh}{\widehat{f^{-1}(0)}}
\newcommand{\fh}{\widehat{f(0)}}

\newcommand{\Xb}{{\bf{X}}}

\newcommand{\Ex}{\mathbb{E}}
\newcommand{\mse}{{\rm MSE}}
\newcommand{\Prob}{\mathbb{P}}
\newcommand{\R}{{\mathbb{R}}}
\newcommand{\Var}{{\mathbb{V}{\rm ar}}}
\newcommand{\Xone}{{X_{(1),N}}}
\newcommand{\bXone}{{\overline{X}_{m_N,s_N}}}

\newtheorem{example}{Example}

\newcommand{\edit}[1]{\textcolor{black}{#1}}

\begin{document}

\title{Minimum Local Distance Density Estimation}
\author{Vikram V. Garg$^1$, Luis Tenorio$^2$, Karen Willcox$^1$}
\affil{vikramvg@mit.edu, ltenorio@mines.edu, kwillcox@mit.edu}
\affil{$^1$Massachusetts Institute of Technology} 
\affil{$^2$Colorado School of Mines}
\date{} 

\maketitle

\begin{abstract} 
We present a local density estimator based on first order statistics. To estimate the density at a point, $x$,
the original sample is
divided into subsets and the average minimum sample distance to $x$ over all
such subsets is used to define the density estimate at $x$. The tuning 
parameter is thus the number of subsets instead of the typical bandwidth of kernel or 
histogram-based density estimators. 
The proposed method is similar to nearest-neighbor density estimators but it provides
smoother estimates.
We derive the asymptotic distribution of
this minimum sample distance statistic to study globally optimal
values for the number and size of the subsets.
Simulations are used to illustrate and compare the
convergence properties of the estimator.
The results show that
the method provides good estimates of a wide variety of
densities without changes of the tuning parameter, and that it
offers competitive convergence performance.

\end{abstract}


\section{Introduction}
Nonparametric density estimation is a classic problem that continues to play an important role
in applied statistics and data analysis. More recently, it has also become a topic of much interest
in computational mathematics, especially in the uncertainty quantification community 
where one is interested in, for example, densities of a large number of coefficients of a 
random function in terms of a fixed set of deterministic functions (e.g., truncated Karhunen-Lo\`eve expansions).
The method we present here was motivated by such applications.

Among the most popular techniques for density estimation are the
histogram \cite{scott1979optimal,scottmv}, kernel~\cite{parzen1962estimation,scottmv,wand} and orthogonal 
series~\cite{efrom,silverman1986density} estimators. For the one-dimensional case,
histogram methods remain in widespread use due to their simplicity and intuitive 
nature, but kernel density estimation has emerged as a method of
choice thanks, in part, to recent adaptive bandwidth-selection methods providing
fast and accurate results~\cite{botev2010kernel}. However, these kernel density estimators can fail to
converge in some cases (e.g., recovering a Cauchy density
with Gaussian kernels)~\cite{buch2005kernel} and can be computationally expensive with large samples
($\orderof{N^2}$, for a sample size $N$). 
Note that histogram estimators are typically implemented using equal-sized bins, 
and nearest-neighbor density estimators
can be roughly thought as histograms whose bins adapt to the local density of the data.
More precisely, let $X_1,\ldots, X_N$ be iid variables from a
distribution with density, $f$, and let $X_{(1)},\ldots,X_{(N)}$ be the corresponding order 
statistics. For any $x$, define $Y_i = |X_i-x|$ and $D_j(x) = Y_{(j)}$.
The $k$-nearest-neighbor estimate of $f$ is defined as (see \cite{silverman1986density} 
for an overview):
\(
\widehat{f}_N(x) = (\,C_N/N\,)/[\,2 D_k(x)\,],
\)
where $C_N$ is a constant that may depend on the sample size.
We may think of $2D_k(x)$ as the width of the bin around $x$. The value of $C_N$ is often chosen as 
$C_N\approx N^{1/2}$ but some effort has been directed towards its 
optimal selection~\cite{fukunaga1973optimization,hall2008choice,li1984consistency},
with some recent work involving the use of order statistics~\cite{kung2012optimal}. 
One of the disadvantages of nearest-neighbor estimators is that their derivative
has discontinuities at the points $(X_{(j)}+X_{(j+k)})/2$, which is caused by the discontinuities
of the derivative of the function $D_k(x)$ at these points. This is clear in 
Figure \ref{fig:smoothdist},
which shows plots of $D_k(x)$ for a sample of size $N=125$ from a Cauchy$(0,1)$ distribution with $k=1$ 
and \edit{$k=round(\sqrt{N})$}. One way to obtain smoother densities
is using a combination of kernel and nearest-neighbor density estimation where the 
nearest-neighbors
technique is used to choose the kernel bandwidth~\cite{silverman1986density}. 
We introduce
an alternative averaging method that improves smoothness and can still be used
to obtain local density estimates.

The main idea of this paper may be summarized as follows: Instead
of using the $k$th nearest-neighbor to provide an estimate of
the density at a point, $x$, we use a subset-average of first order statistics of $|X_i-x|$.
So, the original sample of size $N$ is split into $m$ subsets of size $s$ each; this 
decomposition into subsets allows the control of the asymptotic mean squared error (MSE) 
of the density estimate. Thus, the problem of bandwidth selection is transformed into that of choosing an optimal number
of subsets. This density estimator is naturally parellelizable with 
complexity $\orderof{N^{{1}/{3}}}$ for parallel systems.
 
The rest of this article is organized as follows. In
Section~\ref{sec:theory} we develop the theory that underlies the
estimator and describe asymptotic results.
In Sections~\ref{sec:algo} and \ref{sec:numexp} we describe the actual estimator
and study its performance using numerical experiments.
A variety of densities are used to reveal the strengths and
weaknesses of the estimator. We provide concluding remarks and generalizations 
in Section~\ref{sec:conc}. Proofs and other auxiliary results are collected in Appendix \ref{sec:proofs}. \edit{From here on, when we refer to the size of a sample set being the power of the total number of samples, we assume that it represents a rounded value, for example, $k = \sqrt{N} \Rightarrow k = round(\sqrt{N})$.}


\section{Theoretical framework \label{sec:theory}}

Let $X_1,\ldots,X_N$ be iid random variables from a distribution with invertible CDF,
$F$, and PDF $f$. Our goal is to estimate the value of $f$ at a point, $x_*$,
where $f(x_*)>0$, and where $f$ is either continuous or has a jump discontinuity.
The non-negative random variables $Y_i = |X_i-x_*|$ are iid with PDF:
\(
g(y) = f(y+x_*) + f(x_*-y).
\)
In particular, $f(x_*)=g(0)/2$. Thus, an estimate of $g(0)$ leads to an estimate of
$f(x_*)$. Furthermore, $g$ is more regular than $f$ in a sense described by the following lemma
(its proof and those of the other results in this section are collected in 
Appendix \ref{sec:proofs}). 
\begin{lemma} \label{lemma:gprop} Let $f$ and $g$ be as defined above. Then:
\vspace{-.3cm}
\begin{itemize}
\item[(i)] If $f$ has left and right limits at $x_*$ (i.e., it is either continuous or has a jump discontinuity at $x_*$), then $g$ is continuous at zero.
\vspace{-.2cm}
\item[(ii)] If $f$ has left and right derivatives at $x_*$, then $g$ has a right derivative at zero. Furthermore, if $f$ is differentiable at $x_*$, then $g'(0)=0$.
\end{itemize}
\end{lemma} 
The original question is thus reduced to the following problem: Let $X_1,\ldots,X_N$ be iid
non-negative random variables from a distribution with
invertible CDF, $G$, and PDF $g$.  The goal is to estimate $g(0)>0$
assuming that $g$ is right continuous at zero. The continuity at zero
comes from Lemma \ref{lemma:gprop}(i).  For some 
asymptotic results we also assume that $g$ is right-differentiable with $g'(0)=0$. 
The zero derivative 
is justified by Lemma \ref{lemma:gprop}(ii).  We estimate $g(0)$ using a 
subset-average of first order statistics.

There is a natural connection between density estimation and first order statistics:
If $\Xone$ is the first order statistic of $X_1,\ldots,X_N$, then (under regularity conditions)
$\Ex \Xone \sim Q(1/(N+1))$ as $N\to \infty$,
where $Q = G^{-1}$ is the quantile function, and therefore
$(N+1)\,\Ex \Xone \to 1/g(0)$.
This shows that one should be able to estimate $g(0)$ provided $N$ is large
 and we have a consistent estimate of $\Ex \Xone $. In the next section we provide 
conditions for the limit to be valid and derive a similar limit for the second
moment of $\Xone$; we then define the estimator and study its asymptotics.

\subsection{Limits of first order statistics}

We start by finding a representation of the first two moments of the first-order statistic 
in terms of functions that allow us to determine the limits of the moments as
$N\to \infty$. 

\begin{lemma}\label{lemma:ordmom}
Let $X_1,\ldots,X_N$ be iid non-negative random variables with PDF $g$, invertible CDF $G$ and 
quantile function $Q$. Assume that $g(0)>0$, and define the sequence of functions
\(
 \delta_N(z) = (N+1)(1-z)^N
\)
on $z\in [0,1]$, $N\in \mathbb{N}$. Then:
\vspace{-.3cm}
\begin{itemize}
\item[(i)] 
\begin{eqnarray}
(N+1)\,\Ex \Xone &=& \int_0^1 \frac{\delta_N(z)}{g(Q(z))}\,dz=\int_0^1 Q'(z)\, \delta_N(z)\,dz
\label{eq:EXone}\\
&=&  \frac{1}{g(0)} + \frac{1}{(N+2)}\int_0^1 Q''(z)\,\delta_{N+1}(z)\,dz.\label{eq:EXonev1}
\end{eqnarray}
Furthermore, if $g$ is twice differentiable with $g'(0)=0$, then
\begin{equation} \label{eq:EXonev2}
(N+1)\,\Ex \Xone =  \frac{1}{g(0)} + \frac{1}{(N+2)(N+3)}\int_0^1 Q'''(z)\,\delta_{N+2}(z)\,dz.
\end{equation}
\vspace{-.7cm}
\item[(ii)] If $g$ is differentiable a.e., then
\begin{equation}\label{eq:EXone2}
(N+1)^2\,\Ex[\,\Xone^2\,] = \left(\frac{N+1}{N+2}\right)\int_0^1 (Q^2(z))''\,\delta_{N+1}(z)\,dz.
\end{equation}
\end{itemize}
\end{lemma}
We use the following result to evaluate the limits of the moments as $N\to \infty$.
\begin{prop}\label{prop:convlim}
Let $H$ be a function defined on $[0,1]$ that is continuous at zero, and assume 
there is an integer $m>0$ and a constant $C>0$ such that
\begin{equation}\label{eq:growth}
|H(x)| \leq {C}/{(1-x)^m}
\end{equation}
a.e. on $[0,1]$. Then,\,
\(
\lim_{N\to \infty} \int_0^1 H(x)\,\delta_N(x)\,dx = H(0).
\)
\end{prop}
This proposition allows us to compute the limits of \eqref{eq:EXone}-\eqref{eq:EXone2}
provided the quantile functions satisfy appropriate regularity conditions.
When a function $H$ satisfies \eqref{eq:growth}, we shall say that $H$ 
satisfies a tail condition for some $C>0$ and integer $m>0$. The following
corollary follows from Lemma \ref{lemma:ordmom} and Proposition \ref{prop:convlim}:

\begin{corollary}\label{cor:momlims}
Let $X_1,\ldots,X_N$ be iid non-negative random variables with PDF $g$, invertible CDF $G$ and 
quantile function $Q$. Assume that $g(0)>0$. Then:
\vspace{-.2cm}
\item[(i)] If $g$ is continuous at zero and $Q'$ satisfies a tail condition, then
\begin{equation}\label{eq:limfirst}
\lim_{N\to\infty} (N+1)\,\Ex \Xone = Q'(0) = {1}/{g(0)}.
\end{equation}
If $g$ is differentiable and $Q''$ satisfies a tail condition, then
\begin{equation}\label{eq:limfirstv1}
(N+1)\,\Ex \Xone = {1}/{g(0)} + \orderof{1/N}.
\end{equation}
\item[(ii)] If $g$ is twice differentiable with $g'(0)=0$, $g''$ is continuous at zero and
$Q'''$ satisfies a tail condition, then
\begin{equation}\label{eq:limfirstv2}
(N+1)\,\Ex \Xone = {1}/{g(0)} + \orderof{1/N^2}.
\end{equation}
\item[(iii)] If $g$ is differentiable a.e., $g'$ and $g$ are continuous at zero, and $Q''$ satisfies a tail
condition, then 
\begin{eqnarray}
\lim_{N\to \infty} (N+1)^2\,\Ex[\Xone^2]  &=& 2\,Q'(0)^2 = {2}/{g(0)^2}\label{eq:limsec}\\
\lim_{N\to\infty} \Var\left[\,(N+1)\,\Xone\,\right] & = &  {1}/{g(0)^2}.\label{eq:limvar}
\end{eqnarray}
\end{corollary}

We now provide examples of distributions that satisfy the hypotheses
of Corollary \ref{cor:momlims}. \edit{For these examples, we temporarily return to the notations $X_i$ (iid random variables) and $Y_i= |X_i-x_*|$ used before Lemma~\ref{lemma:gprop}}.

\begin{example}{\rm Let $X_1,\ldots,X_N$ be iid with exponential distribution 
$\mathcal{E}(\lambda)$ and fix $x_*>0$. The PDF, CDF and quantile function of $Y_i$ are,
respectively, 
\begin{eqnarray*}
g(y) &=& 2\lambda\,e^{-\lambda x_*}\cosh(\lambda y)\,I_{y\leq x_*} + \lambda\,e^{-\lambda(x_*+y)}\,I_{y> x_*}\\
G(y) &=& 2e^{-\lambda x_*}\sinh(\lambda y) \,I_{y\leq x_*} + (1-e^{-\lambda(x_*+y)}) \,I_{y> x_*}\\
Q(z) &=& \lambda^{-1}\mathrm{arcsinh} (ze^{\lambda x_*}/2)\,I_{z\leq z_*}-
[\,x_* + \lambda^{-1}\,\log(1-z)\,]\,I_{z>z_*}
\end{eqnarray*} 
for $y\geq 0$, $z\in [0,1)$ and $z_*=1-e^{-2\lambda x_*}$. As expected, $g'(0)=0$. In addition, $Q$ and its derivatives
are continuous at zero. Furthermore, since $|\log(1-z)|\leq z/(1-z)$ on $(0,1)$, we see that $Q$ and its derivatives 
satisfy tail conditions.
}
\end{example}

\begin{example}{\rm Let $X_1,\ldots,X_N$ be iid with Cauchy distribution
and fix $x_*\in\R$. The PDF and CDF of $Y_i$ are:
\begin{eqnarray*}
g(y) &=& \frac{1}{\pi[1+(y+x_*)^2]} + \frac{1}{\pi[1+(x_*-y)^2]}\\
G(y) &=& \arctan(y+x_*)/\pi - \arctan(x_*-y)/\pi.
\end{eqnarray*}
Again, $g'(0)=0$. To verify the conditions on the quantile function, $Q$, note that
\[
Q(z) = -\cot(\pi z) + \cot(\pi z)\sqrt{1 + (1+x_*^2)\tan^2(\pi z)},
\]
in a neighborhood of zero, while for $z$ in a neighborhood of 1, $Q$ is given by
\[
Q(z) = -\cot(\pi z) - \cot(\pi z)\sqrt{1 + (1+x_*^2)\tan^2(\pi z)}.
\]
Since $Q(z)\to 0$ as $z\to 0^+$ and $g$ is smooth, it follows that $Q$ and its derivatives are
continuous at zero. It is easy to see that the tail conditions for $Q'$,
$Q'''$ and $(Q^2)''$ are determined by the tail condition of $\csc(\pi z)$, which
in turn follows from the inequality $|\csc(\pi z)|\leq 1/[\,\pi z(1-z)]$ on
$(0,1)$.
}
\end{example}
 It is also easy to check that the Gaussian and beta distributions satisfy appropriate
tail conditions for Corollary \ref{cor:momlims}.

\subsection{Estimators and their properties}
Let $X_1,\ldots,X_N$ be iid non-negative random variables whose
PDF $g$, CDF $G$ and quantile function $Q$ satisfy appropriate regularity conditions 
for Corollary \ref{cor:momlims}.
We randomly split the sample into $m_N$ independent subsets of size
$s_N$. Both sequences, $(m_N)$ and $(s_N)$, tend to infinity as $N\to \infty$ and satisfy
$m_N s_N = N$. Let $X^{(1)}_{(1),s_N},\ldots,X^{(m_N)}_{(1),s_N}$ be the first-order statistics for each
of the $m_N$ subsets, and let $\bXone$ be their average,
\begin{equation}\label{eq:meanX}
\bXone = \frac{1}{m_N}\sum_{k=1}^{m_N} X^{(k)}_{(1),s_N}.
\end{equation}
The estimators of $1/g(0)$ and $g(0)$ are defined, respectively, as:
\begin{equation}\label{eq:estdefs}
\finvh_{N} = (s_N+1)\bXone,\quad \fh_{N} = 1/\finvh_{N}.
\end{equation}

\begin{prop}\label{prop:mselim}
Let $N$, $m_N$ and $s_N$ be as defined above. Then:
\vspace{-.1cm}
\begin{itemize}
\item[(i)] If $g$ is differentiable a.e., $g'$ and $g$ are continuous at zero, and 
$Q''$ satisfies a tail condition, then 
\begin{equation}\label{eq:mselim}
\lim_{N\to \infty} \mse(\,\finvh_{N}\,) = 0,
\end{equation}
and therefore
\(
\fh_{N} \arrP g(0)
\)
as $N\to \infty$.
\item[(ii)] Let $g$ be twice differentiable with $g'(0)=0$, $g''$ be continuous at zero, and
let $Q'''$ satisfy a tail condition. If $\sqrt{m_N}/s_N\to \infty$ and 
$\sqrt{m_N}/s_N^2\to 0$
as $N\to \infty$, then
\begin{equation}\label{eq:fidistlim}
\sqrt{m_N}\left(\,\finvh_{N}-1/g(0)\,\right) \arrL N(0,1/g(0)^2),
\end{equation}
which leads to
\begin{equation}\label{eq:fdistlim}
\sqrt{m_N}\left(\,\fh_{N}-g(0)\,\right) \arrL N(0,g(0)^2).
\end{equation}
Furthermore, $\mse(\finvh_{N})$ and $\mse(\fh_{N})$ are $\orderof{1/m_N}$. In particular,
\eqref{eq:fidistlim} and \eqref{eq:fdistlim} are satisfied when $s_N = N^\alpha$ and  $m_N = N^{1-\alpha}$
for some $\alpha\in (1/5,1/3)$. This leads to the MSE optimal rate $\orderof{N^{-4/5-\varepsilon}}$ for
any $\varepsilon>0$.
\end{itemize}
\end{prop}

By (ii), we need a balance between the sample size, $s_N$, and the number of
samples, $m_N$: $m_N$ should grow faster than $s_N$ but not much faster.
For comparison, the optimal rate of the MSE is $\orderof{N^{-2/3}}$
for the smoothed histogram, and $\orderof{N^{-4/5}}$ for the kernel density estimator
\cite{dasgupta}.

\subsubsection*{Distance function}

\begin{figure}[!h]
\begin{center}
\includegraphics[keepaspectratio,width=0.5\textwidth]{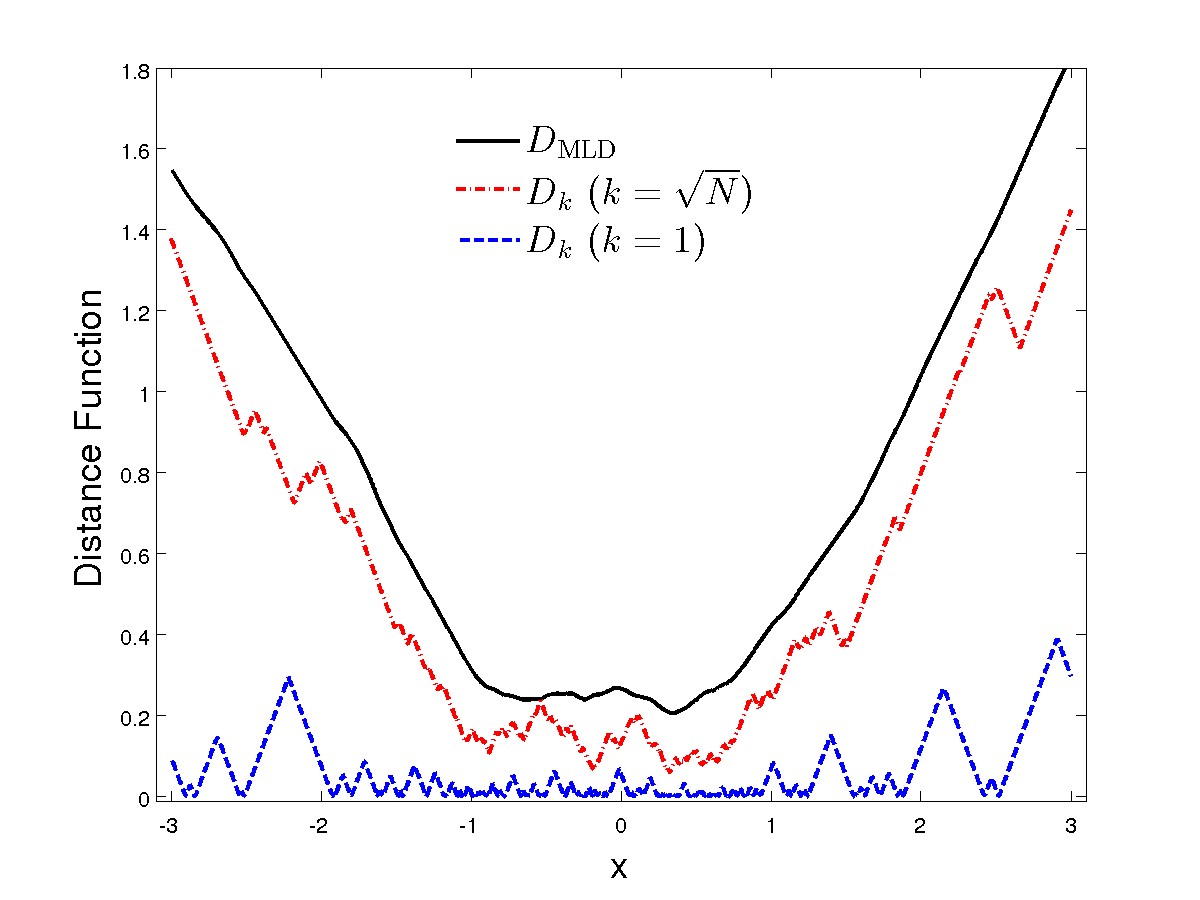}
\caption{Distance function $D_k(x)$ for $k$-nearest-neighbor (for $k=1$ and $k=11\approx \sqrt{N}$) and the distance function 
$D_{\rm MLD}(x)$ \edit{(with $m_N = N^{\frac{2}{3}} = 25$ subsets)} for \edit{125 samples taken from a Cauchy(0,1) distribution.}\label{fig:smoothdist}} 
\end{center}
\end{figure}

We return to the original sample $X_1,\ldots,X_N$ from a density $f$ before the transformation to
$Y_1 = |X_1-x|,\ldots,Y_N = |X_N-x|$. The sample is split into $m_N$ subsets. Let $D_1(x;m)$ be the distance from $x$ 
to its nearest-neighbor in the $m$th subset. The mean $\bXone$ in \eqref{eq:meanX} is the average of $D_1(x;m)$
over all the subsets; we call this average the distance function, $D_{\rm MLD}$, of the MLD density estimator. That is,
\[
D_{\rm MLD}(x) = \bXone=\frac{1}{m_N}\sum_{m=1}^{m_N} D_1(x;m).
\]
The estimators in \eqref{eq:estdefs} can then be written in terms of $D_{\rm MLD}(x)$.
This distance function tends to be smoother than the usual distance function used by $k$-nearest-neighbor density 
estimators. For example, Figure \ref{fig:smoothdist} shows the different distance functions $D_{\rm MLD}(x)$, $D_1(x)$
and $D_k(x)$ (the latter as defined in the introduction) for a sample of $N=125$ variables
from a Cauchy$(0,1)$. Note that $D_{\rm MLD}$ is an average of first-order statistics for samples of size
$s_N$, while $D_1$ is a first-order statistic for a samples of size $N$,  so $D_{\rm MLD}>D_1$. On the
other hand, $D_{\sqrt{N}}$ is a $N^{1/2}$th-order statistic based on a sample of size $N$; hence the order
 $D_{\rm MLD}>D_{\sqrt{N}}>D_1$.

\section{Minimum local distance density estimator \label{sec:algo}}
We now describe the 
local distance density estimator (MLD-DE). The inputs are: a sample, a set of points where 
the density is to be estimated and the parameter $\alpha$ whose default is set 
to $\alpha=1/3$.
The basic steps to obtain the density estimate at a point $x$ are: (1) Start with a sample 
of $N$ iid variables from the unknown density, $f$; (2) Randomly split the sample into 
$m_N = N^{1-\alpha}$ disjoint subsets of size $s_N = N^{\alpha}$ each; 
(3) Find the nearest sample distance to $x$ in each subset; (4) Compute the density estimate by 
inverting the average nearest distance across the subsets and scaling it (see~Eq.\eqref{eq:estdefs}).
This is summarized in Algorithm \ref{alg:MLD}.
\begin{algorithm}[!h]
  \caption{\label{alg:MLD} Returns density estimates at the points \edit{of evaluation $\{x_{l}\}_{l=1}^{M}$} given the sample
 $X_1,\ldots,X_N$ from the unknown density $f$.}
  \label{alg:nnj}
  \begin{algorithmic}[1]
    \State $m_N \leftarrow$ round($N^{1-a}$)
    \State $s_N \leftarrow$ round(${N}/{m_N}$)
    \State Create an $s_N \times m_N$ matrix $M_{ij}$ with the $m_N$ subsets 
with $s_N$ variables each
    \State Create a vector $\widehat{f}=(\widehat{f}_\ell)$ to hold the density estimates 
at the points $\{x_{l}\}_{l=1}^{M}$ 
    \For{$l = 1 \to M$} 
    \For{$k = 1 \to s_N$}     
    \State Find the nearest distance $d_{lk}$ to the current point $x_\ell$ within the $k$th subset
    \EndFor
    \State Compute the subset average of distances to $x_\ell$: $d_\ell = (1/m_N) 
\sum\limits_{k=1}^{m_N}d_{\ell k}$  
    \State Compute the density estimate at $x_l$: $\widehat{f}_\ell = {1}/{2 d_\ell}$ 
    \EndFor \\
\Return $\widehat{f}$
   \end{algorithmic}
\end{algorithm}
Note that for each of the $M$ points where the density is to be estimated, the algorithm loops over $N^{1-\alpha}$ 
subsets, and within each it does a nearest-neighbor search over $N^{\alpha}$ points. The computational complexity 
is therefore $\orderof{M N^{1-\alpha} N^{\alpha}} = \orderof{MN}$, which is of the same order as the $\orderof{N^2}$ 
complexity of kernel density estimators~\cite{raykar2010fast} when $M \sim N$. However, MLD-DE
displays multiple levels of parallelism. The first level is the highly 
parallelizable evaluation of the density at the $M$ specified points. The second level arises 
from the the nearest-neighbor distances that can be computed independently in each subset. 
Thus, for parallel systems the effective computational 
complexity of the algorithm is $\orderof{MN^{\alpha}}$, which is the same as that of
histogram methods if $\alpha = {1}/{3}$. 

\section{Numerical examples \label{sec:numexp}}
An extensive suite of numerical experiments was used to test the MLD-DE method.
We now summarize the results to show that they
are consistent with the theory derived in Section~\ref{sec:theory}, and
illustrate some salient features of the estimator. 
We also compare MLD-DE to the
adaptive kernel density estimator (KDE) introduced by Botev et
al.~\cite{botev2010kernel} and to the histogram method based on Scott's normal 
reference rule~\cite{scott1979optimal}.

We first discuss experiments for density estimation at
a fixed point and show the effects of changing the number of subsets for a
fixed sample size. We then estimate the integrated mean square error for various densities, 
and compare the convergence of MLD-DE to that of other density estimators. Next, we present numerical experiments that show
the spatial variation of the bias and variance of MLD-DE,
and relate them to the theory derived in Section~\ref{sec:theory}. Finally, 
we check the impact of changing the tuning parameter $\alpha$
(see Proposition~\ref{prop:mselim}).
\subsection{Pointwise estimation of a density}
\begin{figure}[!h]
\begin{center}
\includegraphics[keepaspectratio,width=0.45\textwidth]{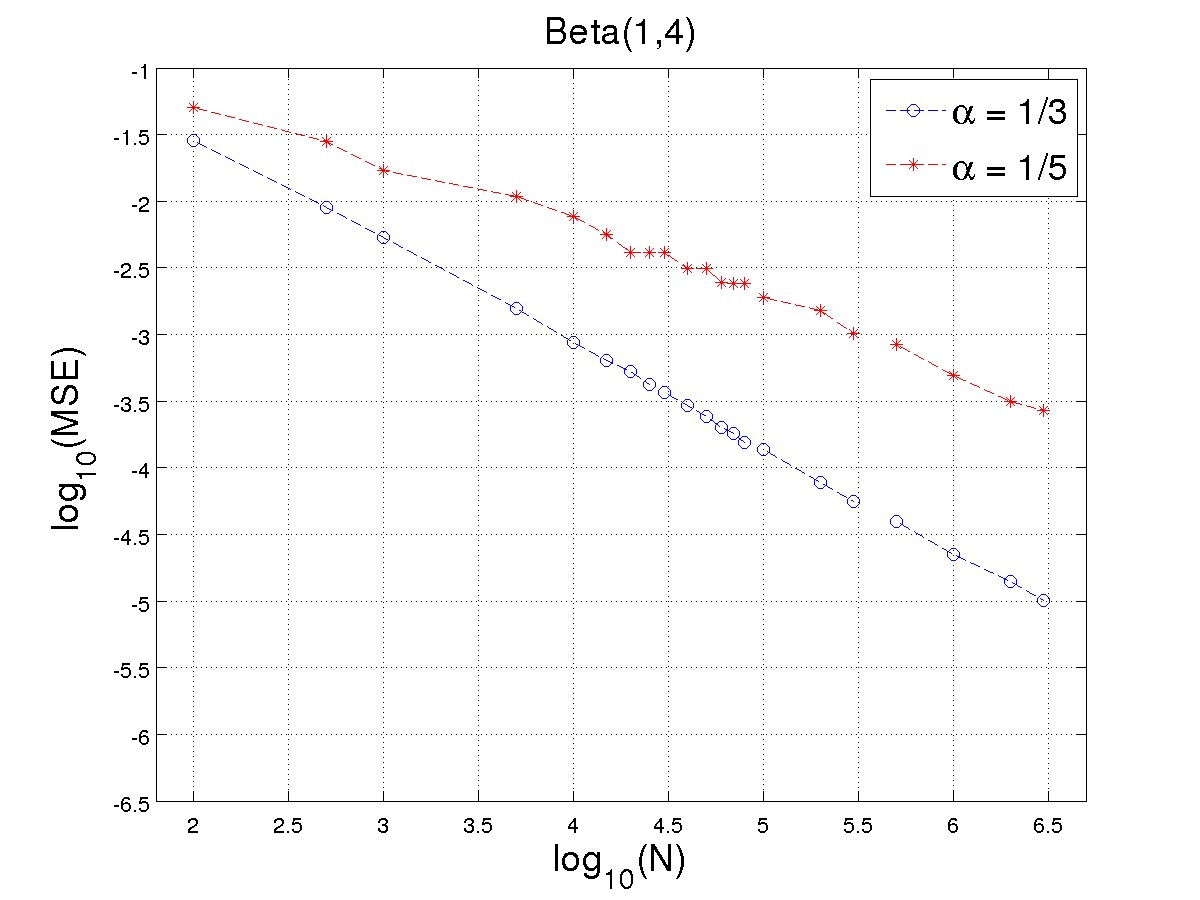}\hspace{.2cm}
\includegraphics[keepaspectratio,width=0.45\textwidth]{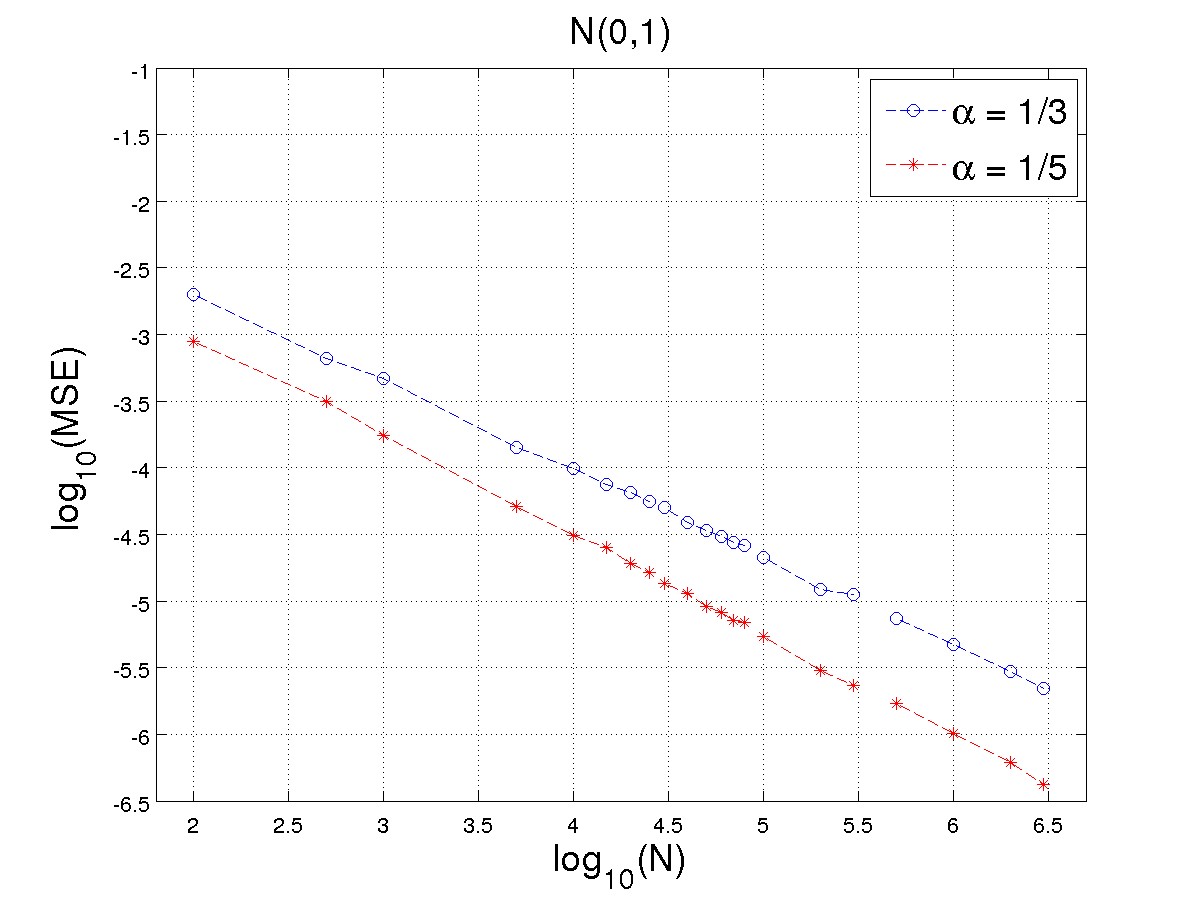}
\caption{Convergence plots of the density estimates at $x=1/2$ for the distribution beta$(1,4)$ (left),
and at $x=1$ for $N(0,1)$ (right).\label{fig:pointwise_conv}} 
\end{center}
\end{figure}

We use MLD-DE to estimate values of the beta$(1,4)$ and
$N(0,1)$ densities at a single point and analyze its convergence
performance. Starting with a sample size $N=100$, $N$ was
progressively increased to three million.  For each $N$,
1000 trials were performed to estimate the MSE of the density
estimate. The parameter $\alpha$ was also
changed; it was set to ${1}/{3}$ for one set of experiments
anticipating a bias of $\orderof{{1}/{N}}$, and to ${1}/{5}$ for
another set, anticipating a bias of $\orderof{{1}/{N^2}}$. The results
are shown in Figure~\ref{fig:pointwise_conv}.
We see the contrasting convergence behavior for the beta$(1,4)$ and $N(0,1)$ distributions. 
For the former, the convergence is faster when $\alpha =
{1}/{3}$, while for the Gaussian it is
faster with $\alpha= {1}/{5}$. We recall from Section \ref{sec:theory}
that the asymptotic bias of the density estimate at a point is
$\orderof{{1}/{N^2}}$. However, reaching the asymptotic regime depends
on the convergence of $\int_{0}^{1} Q''(z) \, \delta_N(z)
\, dz $ to zero, which can be quite slow, depending on the behavior of
the density at the chosen point. Hence, the effective bias in
simulations can be $\orderof{{1}/{N}}$. The numerical experiments thus
indicate that the quantile function derivative of the Gaussian decays
to zero much faster than that of the beta distribution, and hence the
optimal value of $\alpha$ for $N(0,1)$ is ${1}/{5}$, while that for beta$(1,4)$
is ${1}/{3}$. However, in either case the order of the decay in the figure is close to
$N^{-3/4}$.
%
%
\subsection{$L^2$-convergence}
We now summarize simulation results regarding the $L^2$-error (i.e., integrated MSE) of estimates of a beta$(1,4)$, 
a Gaussian mixture and the Cauchy$(0,1)$ density. The Gaussian mixture used is (see ~\cite{wasserman2006all}):
\(
\label{eq:gm_pdf}
0.5\, N(0,1) + 0.1 \sum_{i=0}^{4} N(\,i/2-1,\,1/100^2\,).
\)
For comparison, these densities were estimated using MLD-DE,
the Scott's rule-based histogram, and the adaptive
KDE proposed by~\cite{botev2010kernel}. Both, the Scott's rule-based
histograms and KDE method fail to recover the Cauchy$(0,1)$
density. For the histogram method, this limitation was overcome using an interquartile range
(IQR) based approach for the Cauchy density that uses a bandwidth, $h_N$, based on the 
Freedman-Diaconis rule~\cite{freedman1981histogram}:
\begin{equation}
\label{eq:iqr_bwith}
h_N =2\, N^{-1/3}\,\mathrm{IQR}_N,
\end{equation}
where IQR$_N$ is the sample interquartile range for a sample of size $N$. For the KDE, 
there is no clear method that
enables us to estimate a Cauchy density, thus KDE  was only used for the Gaussian mixture and beta densities.
\begin{figure}[!h]
\begin{center}
\subfigure[Beta(1,4) distribution]
{
\includegraphics[width=0.35\textwidth]{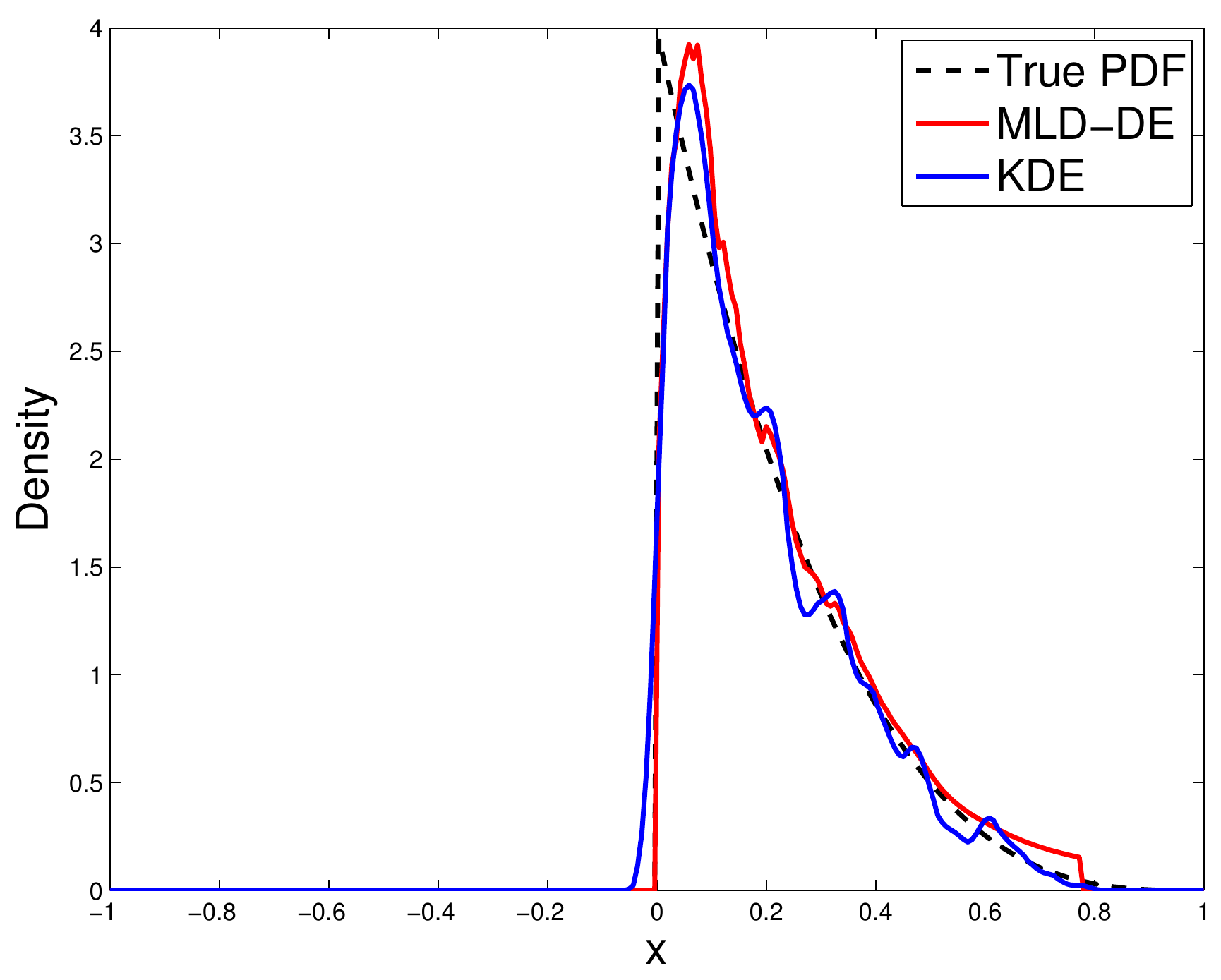}
\label{fig:beta_comparison}
}
\subfigure[Gaussian Mixture]
{
\includegraphics[width=0.35\textwidth]{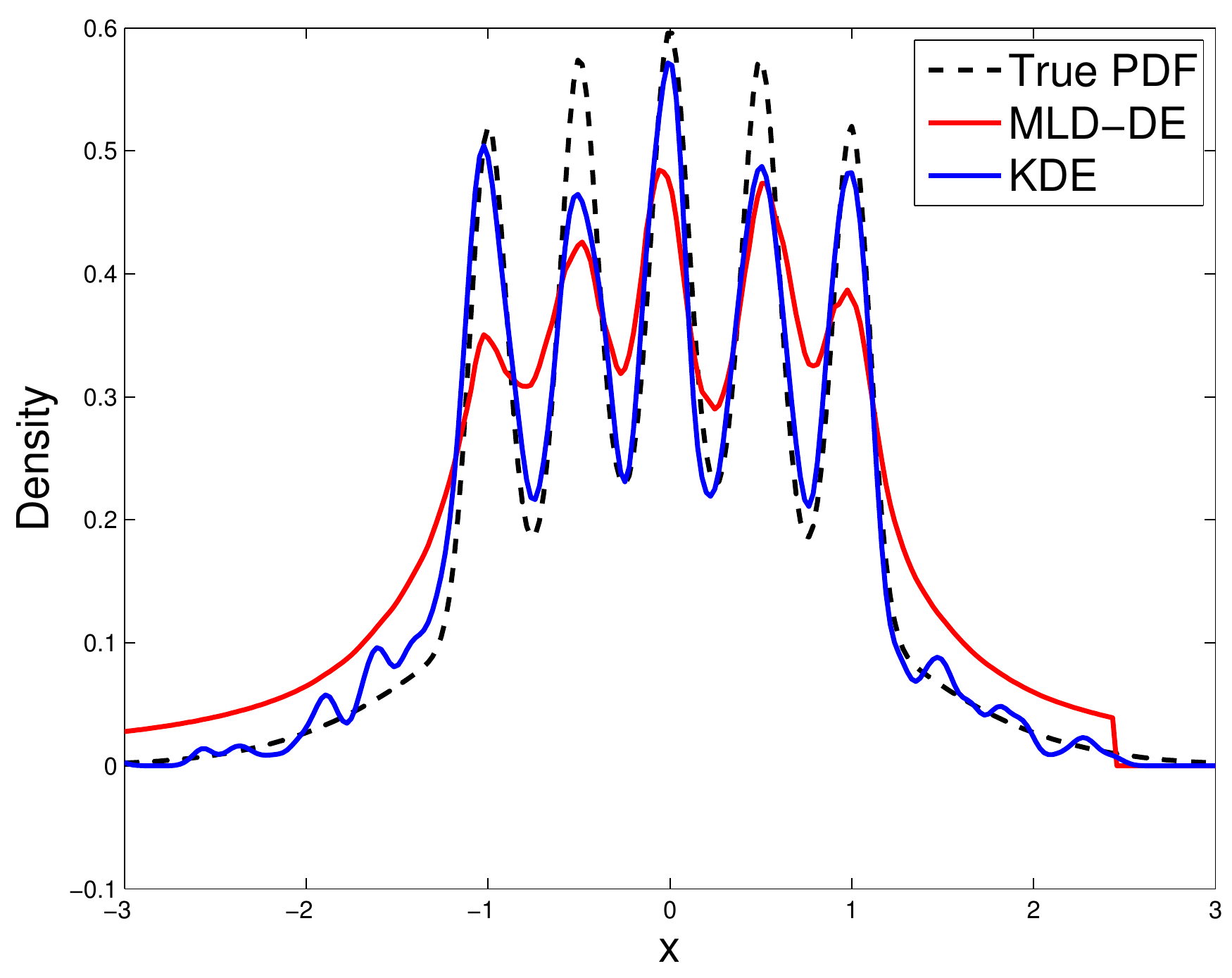}
\label{fig:gm_comparison}
}\\
\subfigure[Gaussian Mixture: optimal $\alpha$]
{
\includegraphics[width=0.35\textwidth]{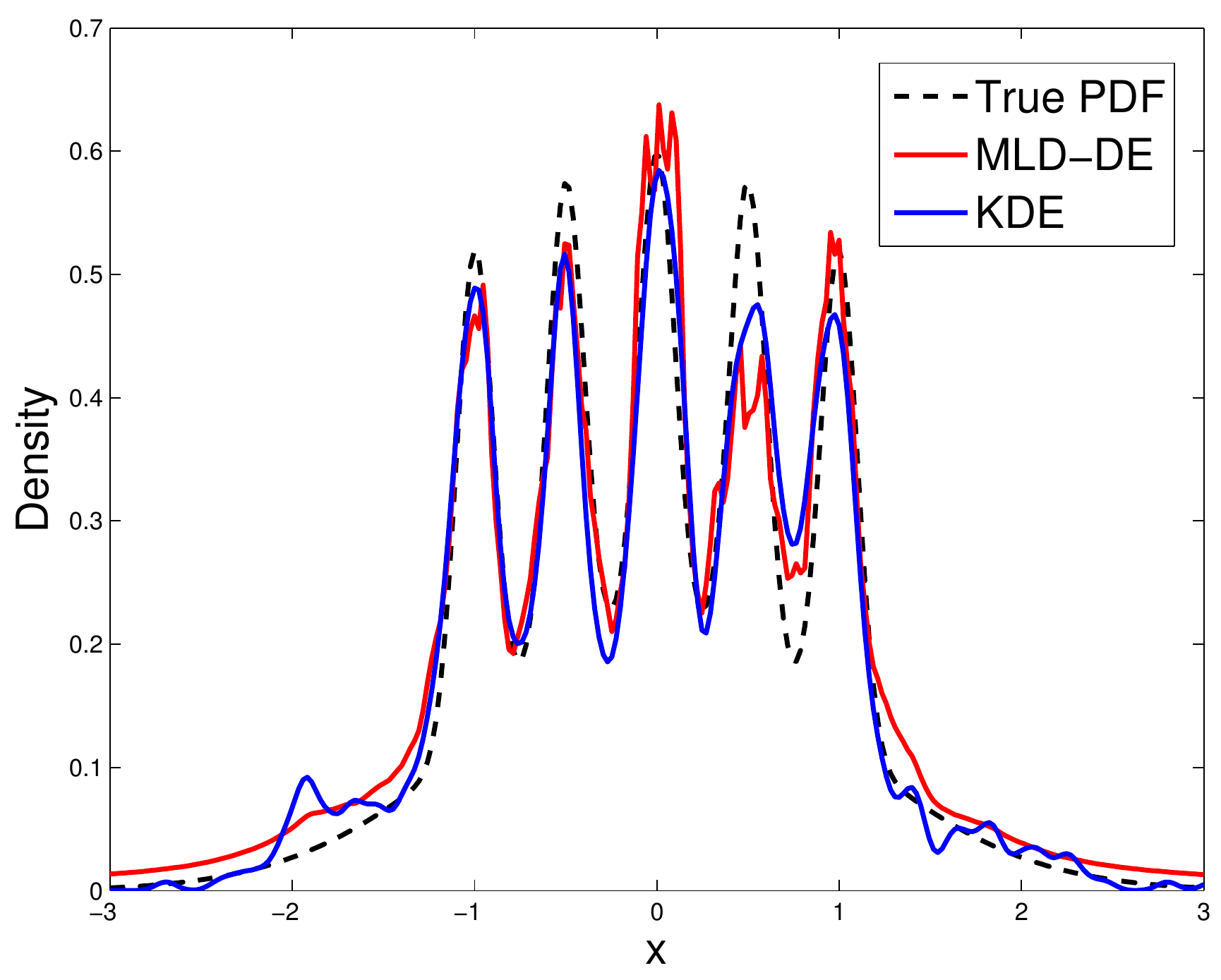}
\label{fig:gm_comparison_optimal_a}
}
\subfigure[Cauchy(0,1) distribution]
{
\includegraphics[width=0.35\textwidth]{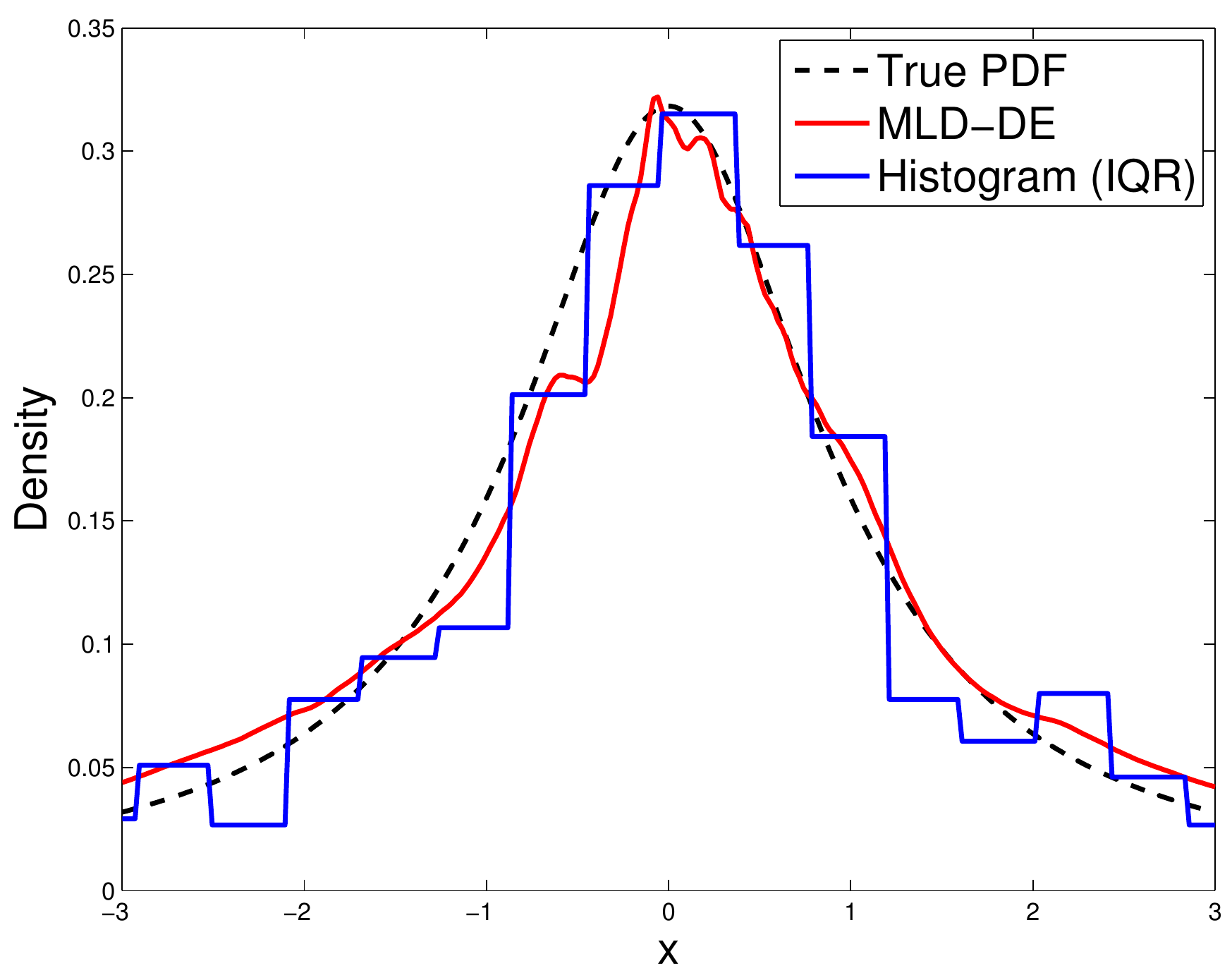}
\label{fig:cauchy_comparison}
}
\caption{Density estimates using MLD-DE, KDE and histogram approaches for the beta$(1,4)$, 
Gaussian mixture and Cauchy$(0,1)$ distributions.}
\label{fig:comparison}
\end{center}
\end{figure}

For the MLD-DE and histogram-based estimators, estimates were obtained for 256 points in 
specified intervals. The interval used for each distribution is 
shown in the figures as the range over which the densities are plotted. 
Once the pointwise density 
estimates were calculated, interpolated density estimates were obtained using nearest-neighbor 
interpolation. For example, Figure~\ref{fig:comparison} shows density estimates from a single 
sample using $\alpha=1/3$ for the beta (Figure~\ref{fig:beta_comparison}), Gaussian Mixture 
(Figure~\ref{fig:gm_comparison}) and Cauchy (Figure~\ref{fig:cauchy_comparison}), and 
with an optimal $\alpha$
for the Gaussian mixture 
(Figure~\ref{fig:gm_comparison_optimal_a}) obtained by simulation.

\begin{figure}[!h]
\begin{center}
\subfigure[Beta(1,4) distribution]
{
\includegraphics[width=0.35\textwidth]{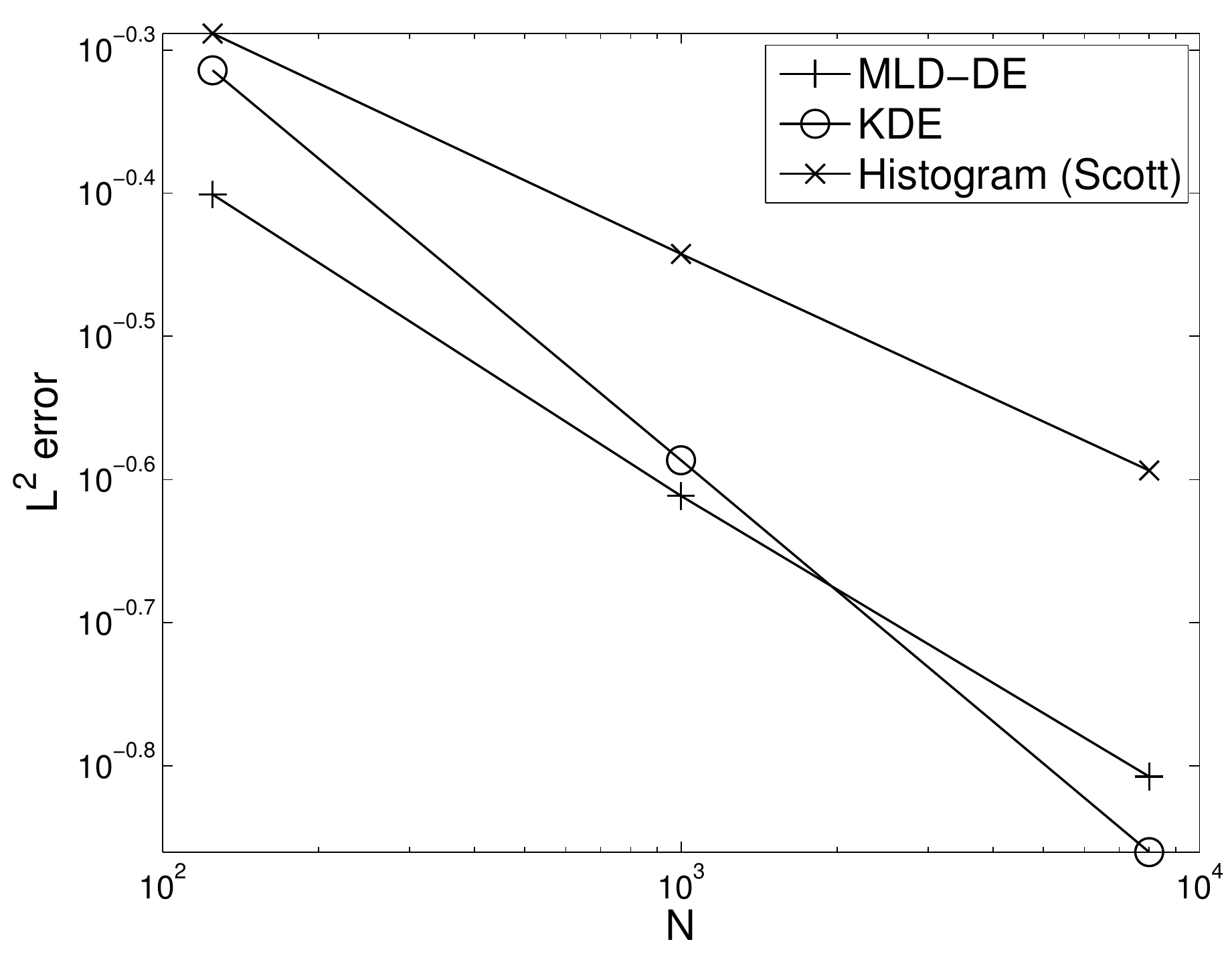}
\label{fig:beta_L2}}
\subfigure[Gaussian Mixture]
{
\includegraphics[width=0.35\textwidth]{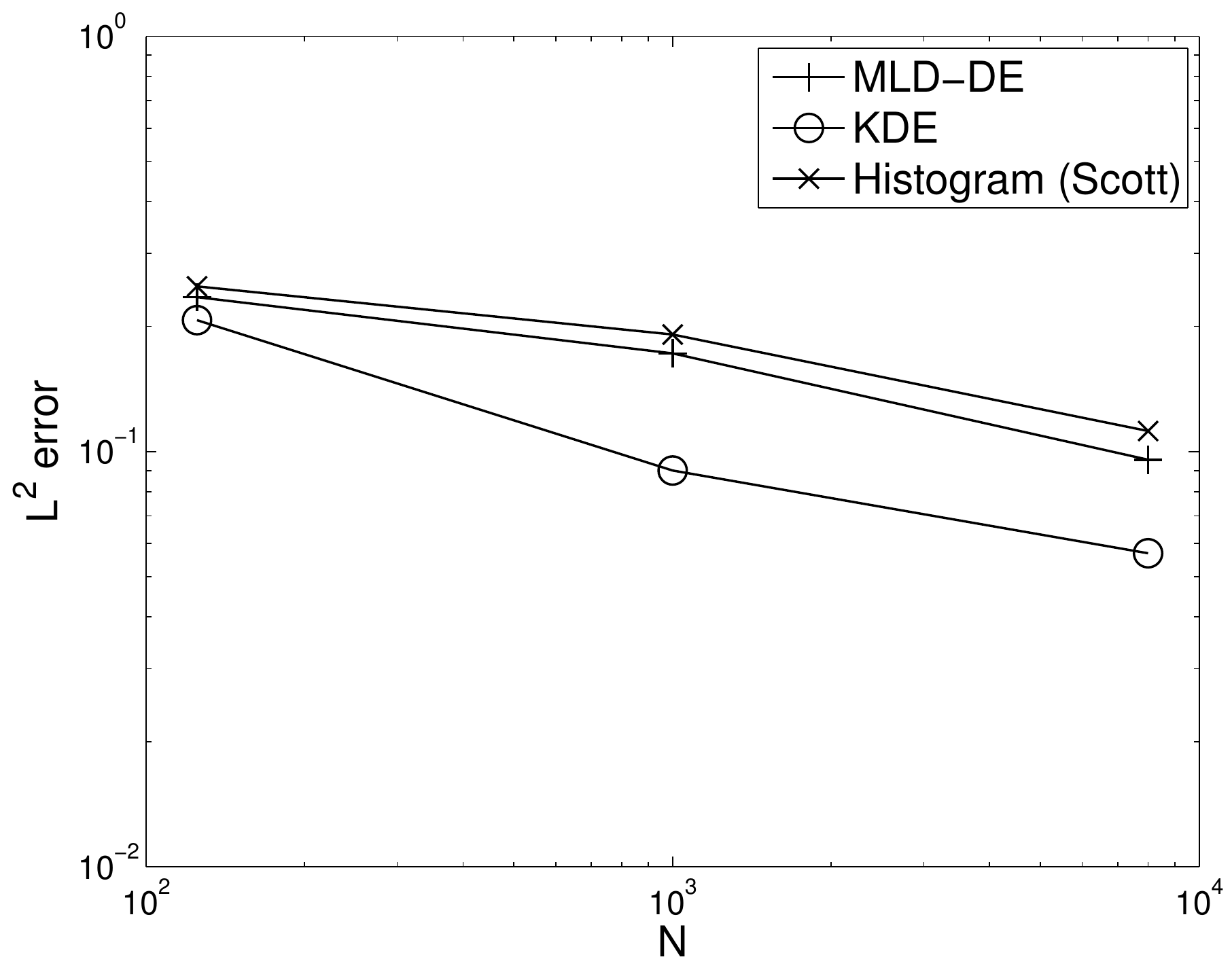}
\label{fig:gmixture_L2}}
\subfigure[Gaussian Mixture: optimal $\alpha$]
{
\includegraphics[width=0.35\textwidth]{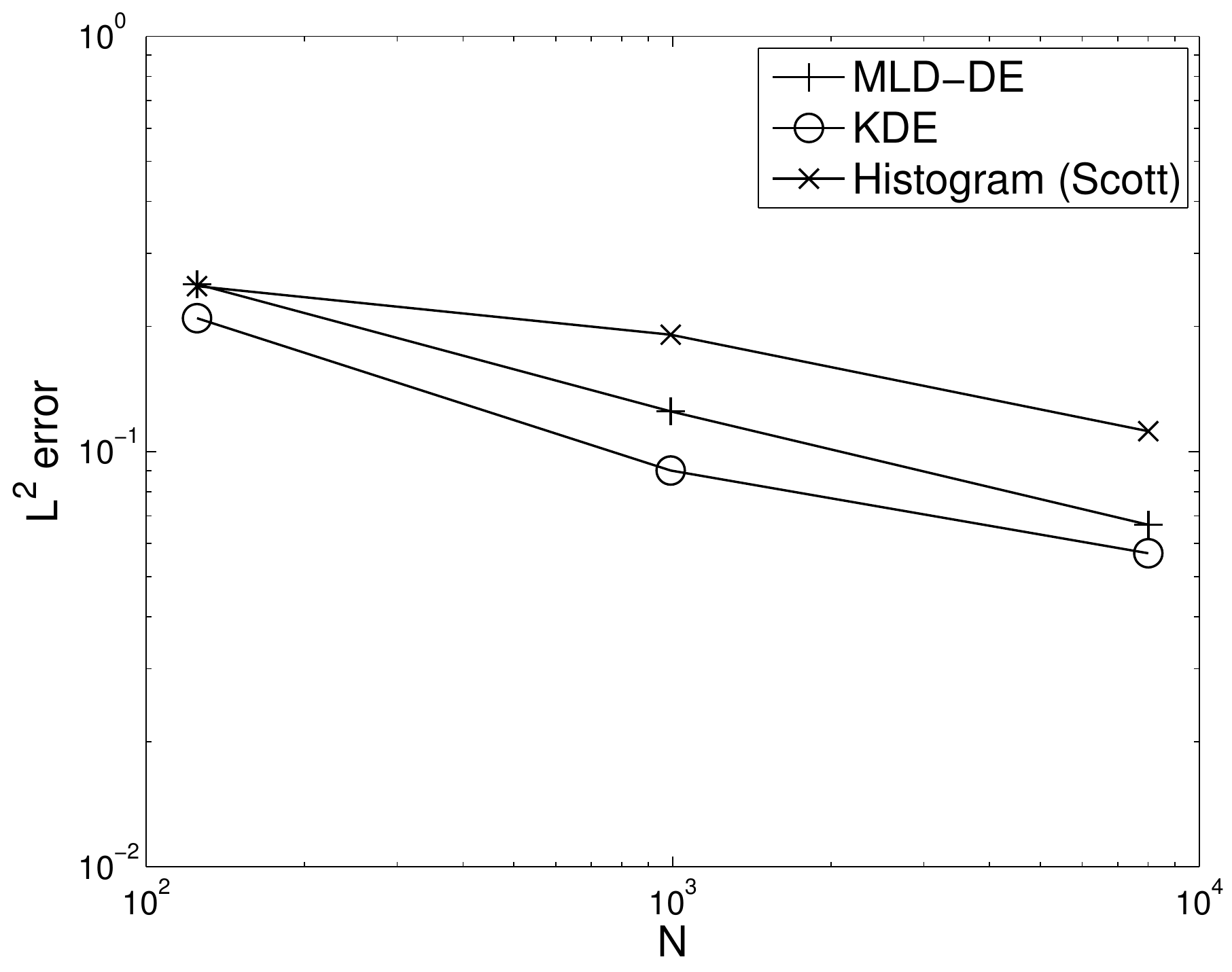}
\label{fig:gmixture_L2_opt_a}}
\subfigure[Cauchy(0,1) density]
{
\includegraphics[width=0.35\textwidth]{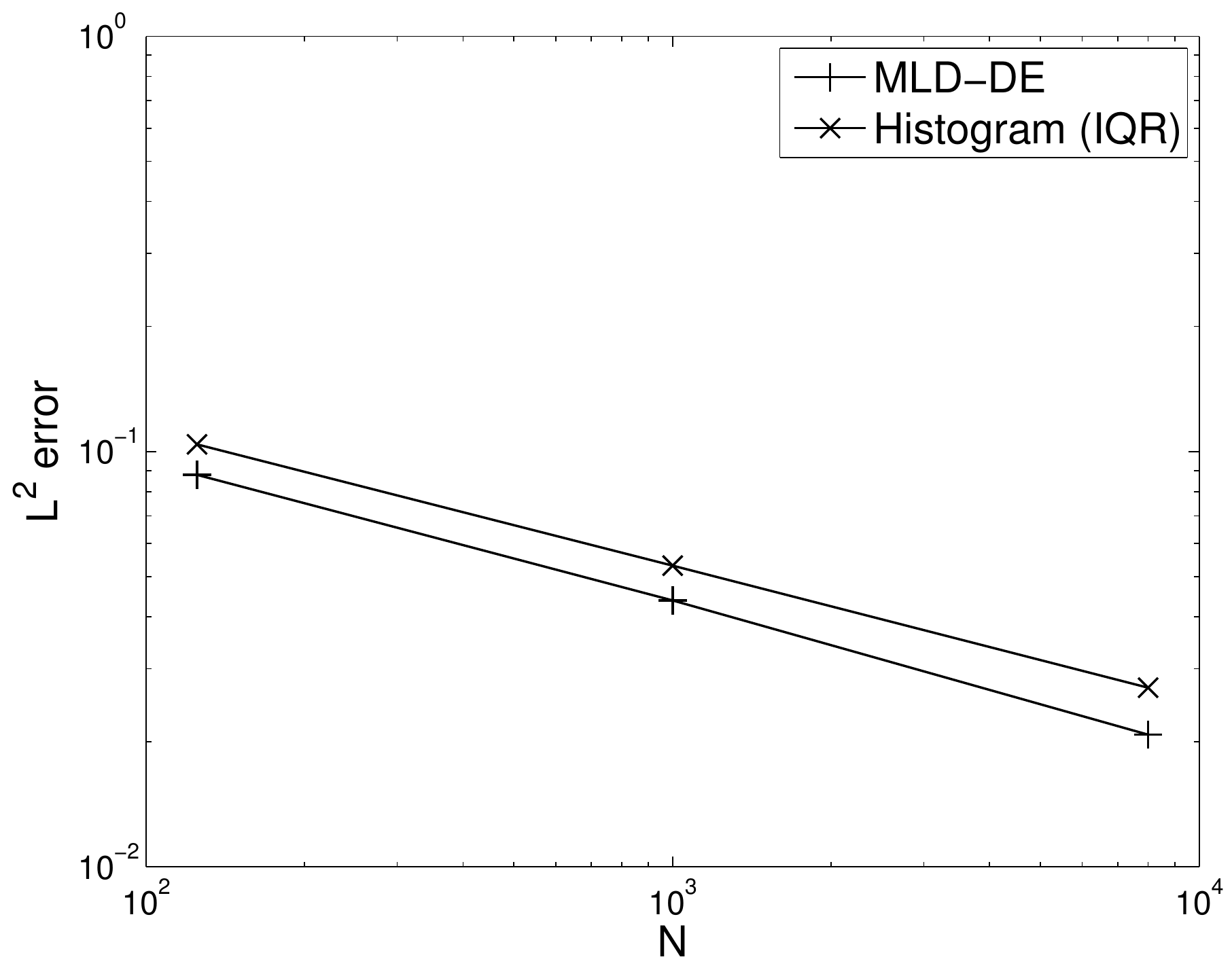}
\label{fig:cauchy_L2}}
\caption{$L^{2}$-convergence plots for various densities.}
\label{fig:L2_conv} 
\end{center}
\end{figure}

The sample size was again increased progressively starting with
$N=125$ up to a maximum sample size $N=8000$. The MSE was
calculated at every point of estimation, and then numerically
integrated to obtain an estimate of the $L^2$-error. A total of 1000 trials
were performed at each sample size to obtain the expected $L^2$-error for such
sample size. Figure~\ref{fig:L2_conv} shows the convergence plots
obtained for the three densities using the various density estimation
methods (the error bars are the size of the plotting symbols). We see that the
performance of MLD-DE is comparable to
that of the histogram method for the beta and Gaussian mixture
densities, and KDE performs better with both these densities.
For the Cauchy density, both the histogram based on
Scott's rule and the KDE approach fail to converge. This is because
Scott's rule requires a finite second moment, whereas the kernel used
in the KDE estimator is a Gaussian kernel, which has finite moments. But
MLD-DE produces convergent estimates of the Cauchy density
without any need to change the parameters from those used with the
other densities. Furthermore, it also performs better than the
IQR-based histogram, which is designed to be less sensitive to
outliers in the data. Thus, MLD-DE provides a robust alternative to
the histogram and kernel density estimation methods, while offering
competitive convergence performance.

\subsection{Spatial variation of the pointwise error}
\begin{figure}[!h]
\begin{center}
\subfigure[Beta(1,4) distribution]
{
\includegraphics[width=0.35\textwidth]{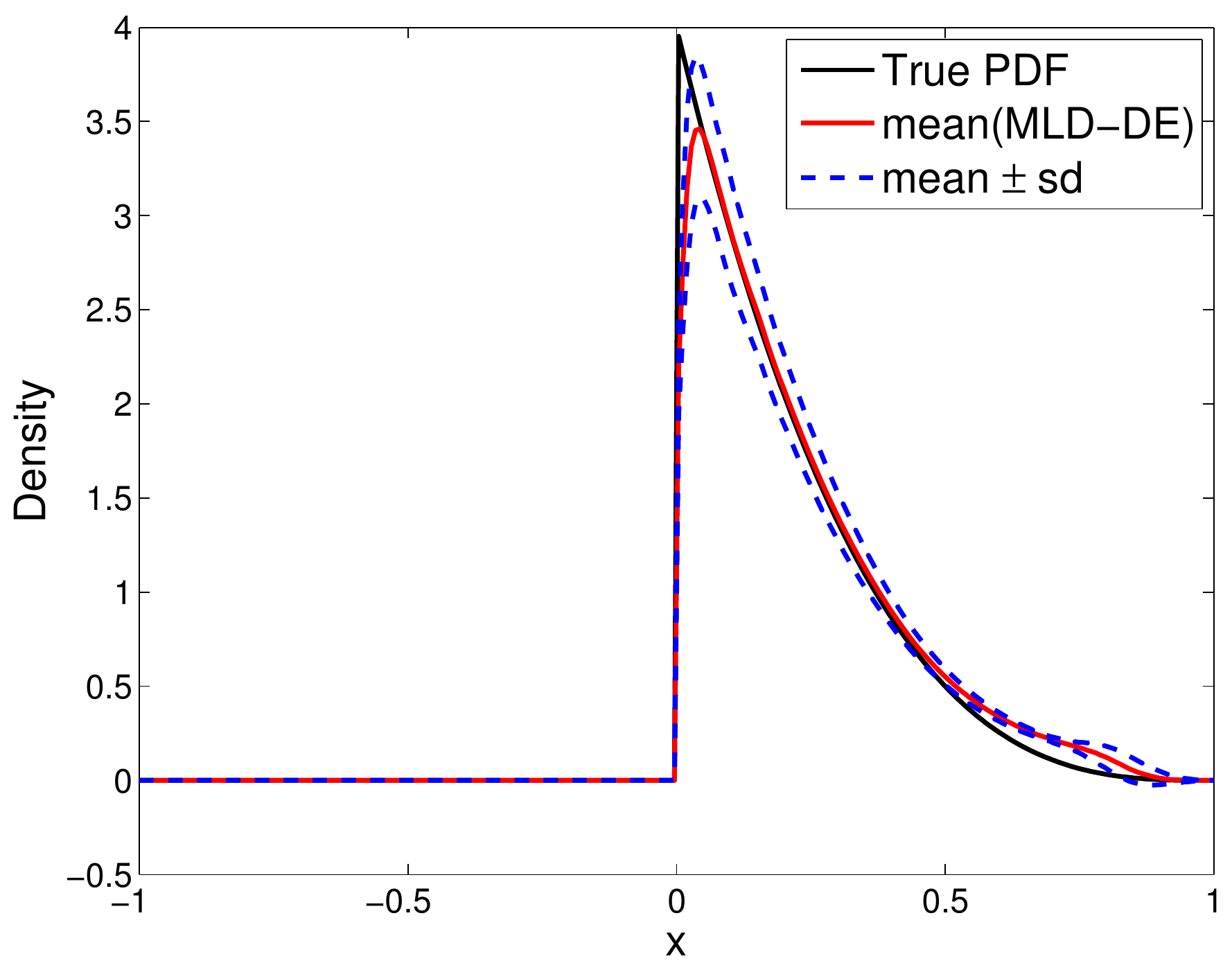}
\label{fig:beta_std}}
\subfigure[Gaussian Mixture]
{
\includegraphics[width=0.35\textwidth]{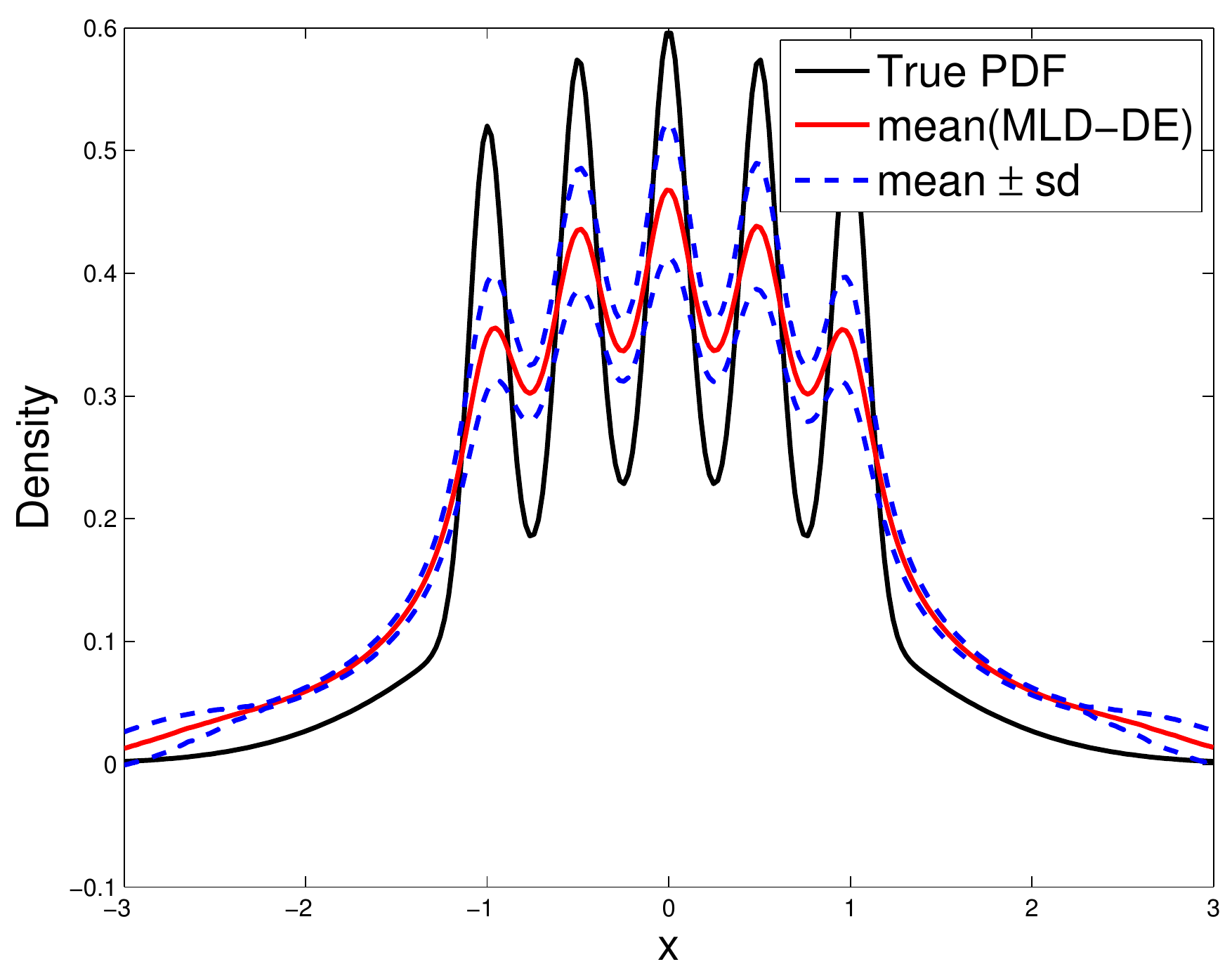}
\label{fig:gmixture_std}}
\subfigure[Gaussian mixture: optimal $\alpha$]
{
\includegraphics[width=0.35\textwidth]{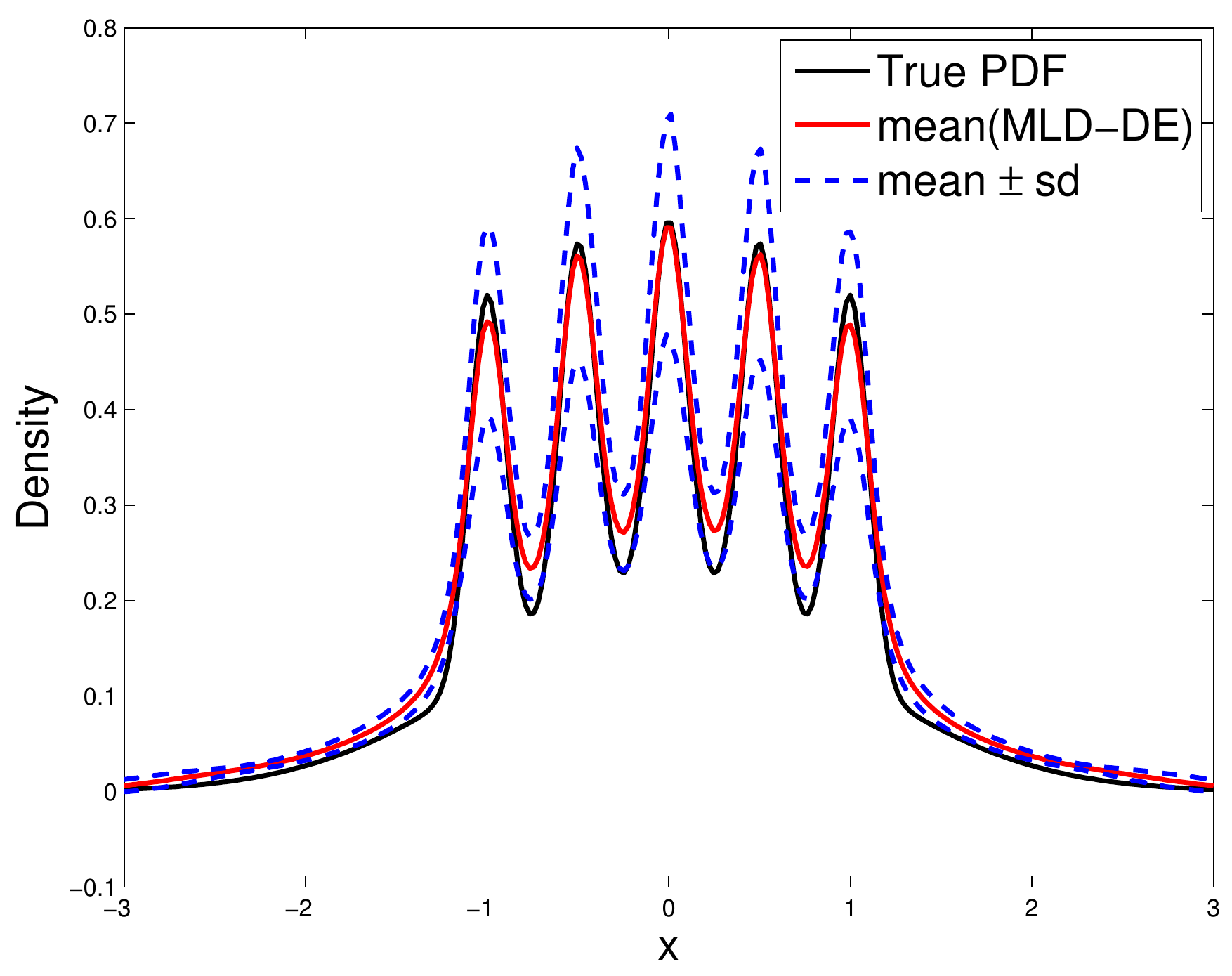}
\label{fig:gmixture_std_opt_a}}
\subfigure[Cauchy(0,1) distribution]
{
\includegraphics[width=0.35\textwidth]{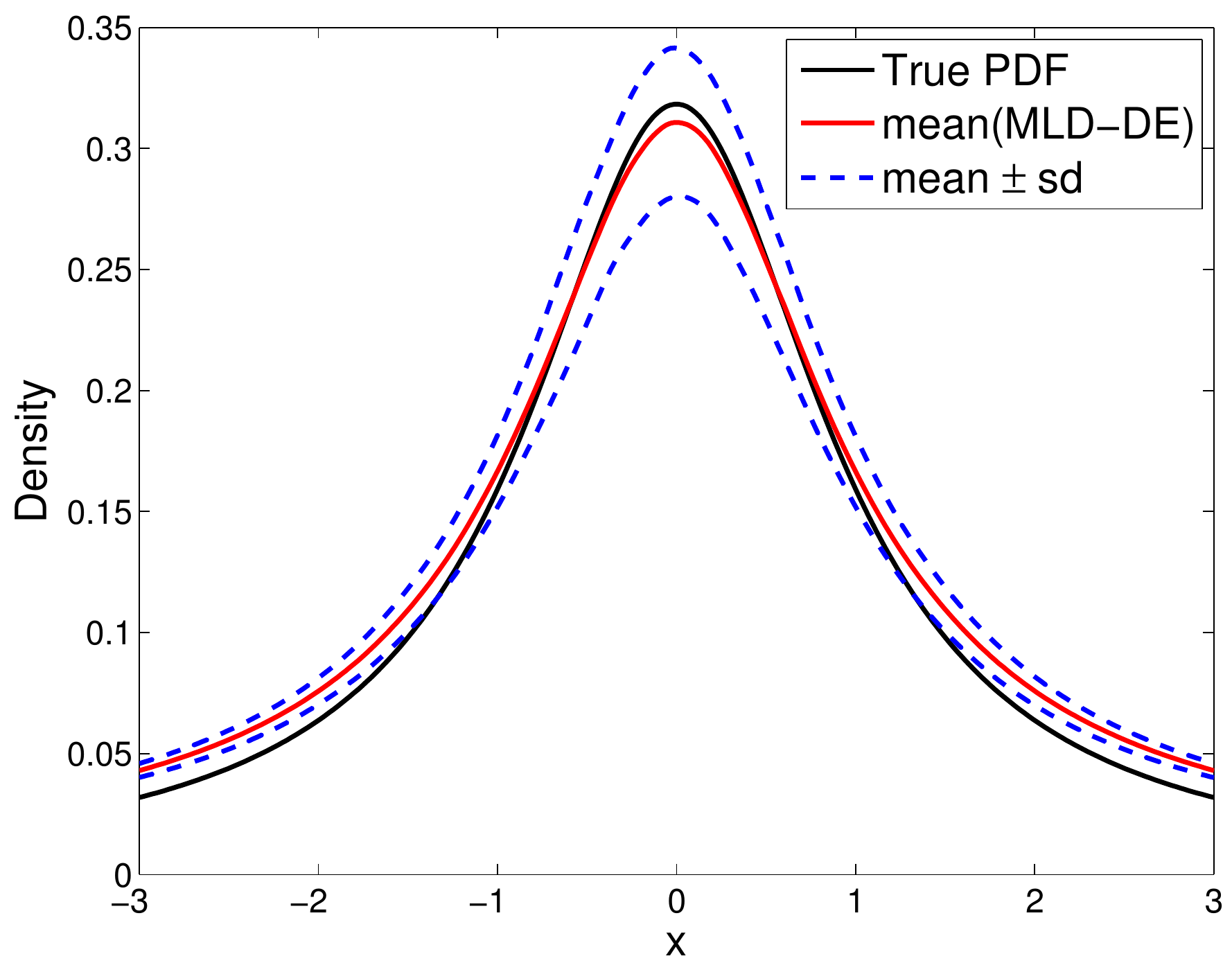}
\label{fig:cauchy_std}}
\caption{Pointwise mean and variance of the MLD-DE estimates for various densities.}
\label{fig:std_bias} 
\end{center}
\end{figure}

\begin{figure}[!h]
\begin{center}
\subfigure[Beta(1,4) distribution]
{
\includegraphics[width=0.35\textwidth]{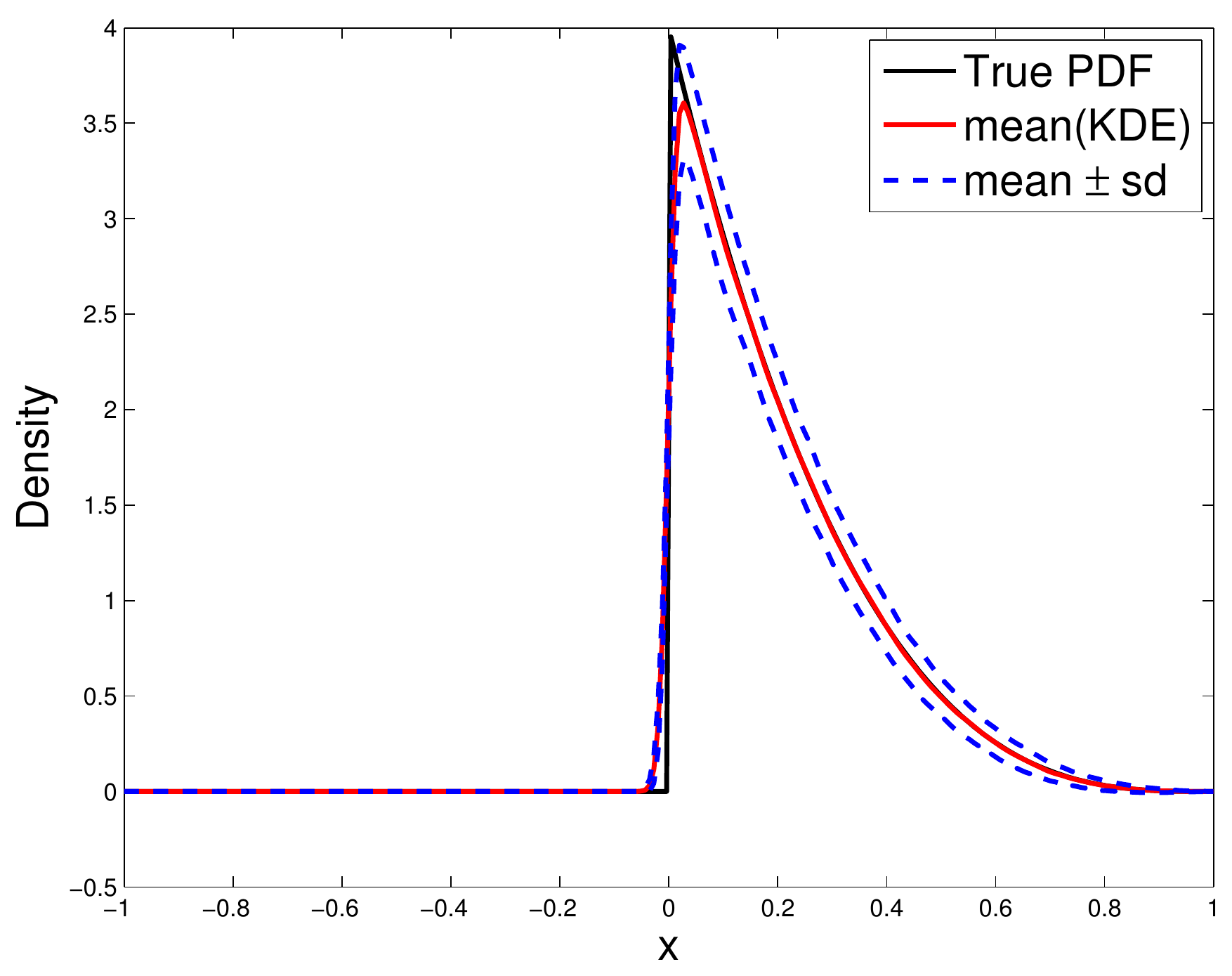}
\label{fig:beta_std_kde}}
\subfigure[Gaussian Mixture]
{
\includegraphics[width=0.35\textwidth]{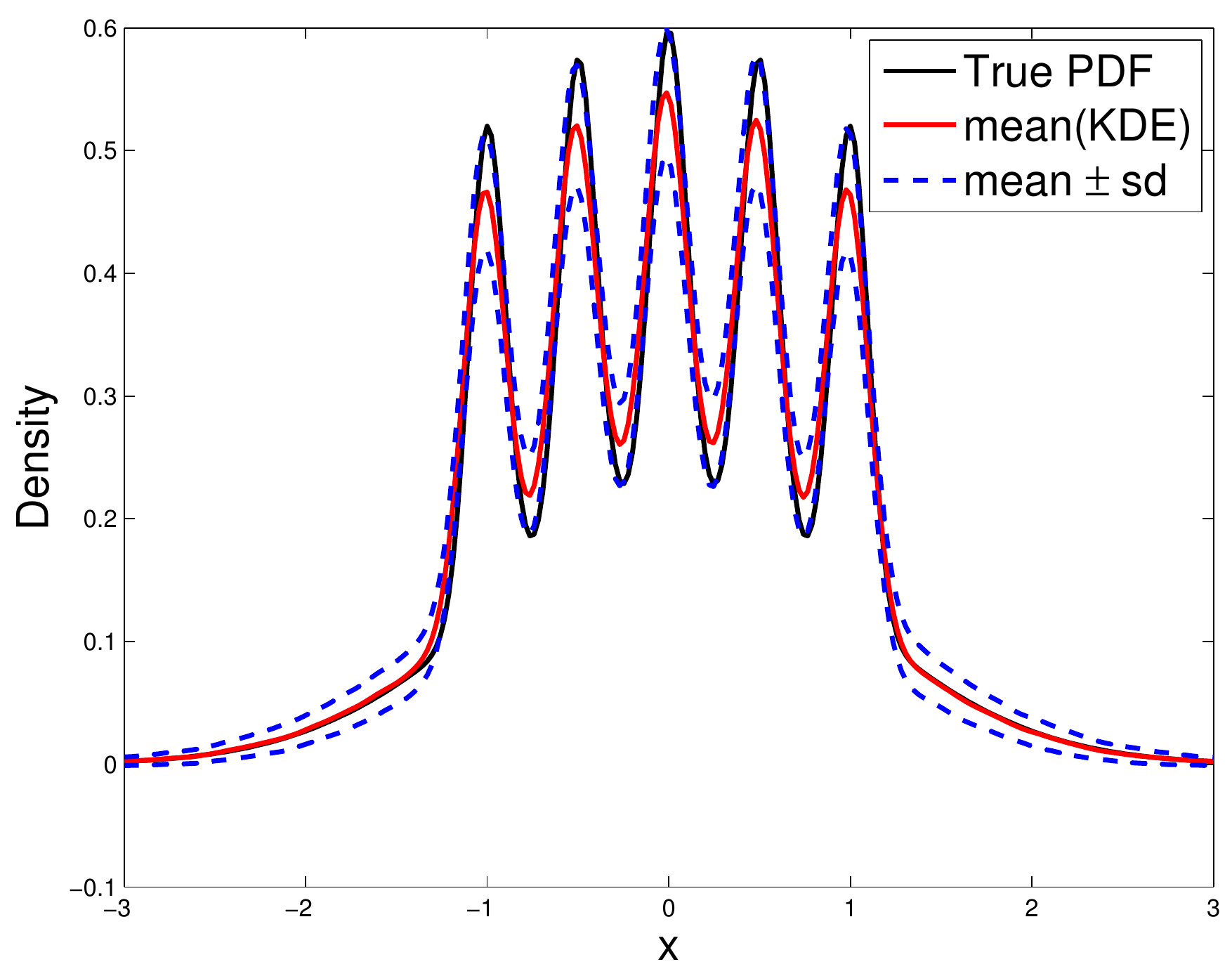}
\label{fig:gmixture_std_kde}}
\subfigure[Cauchy(0,1) distribution]
{
\includegraphics[width=0.35\textwidth]{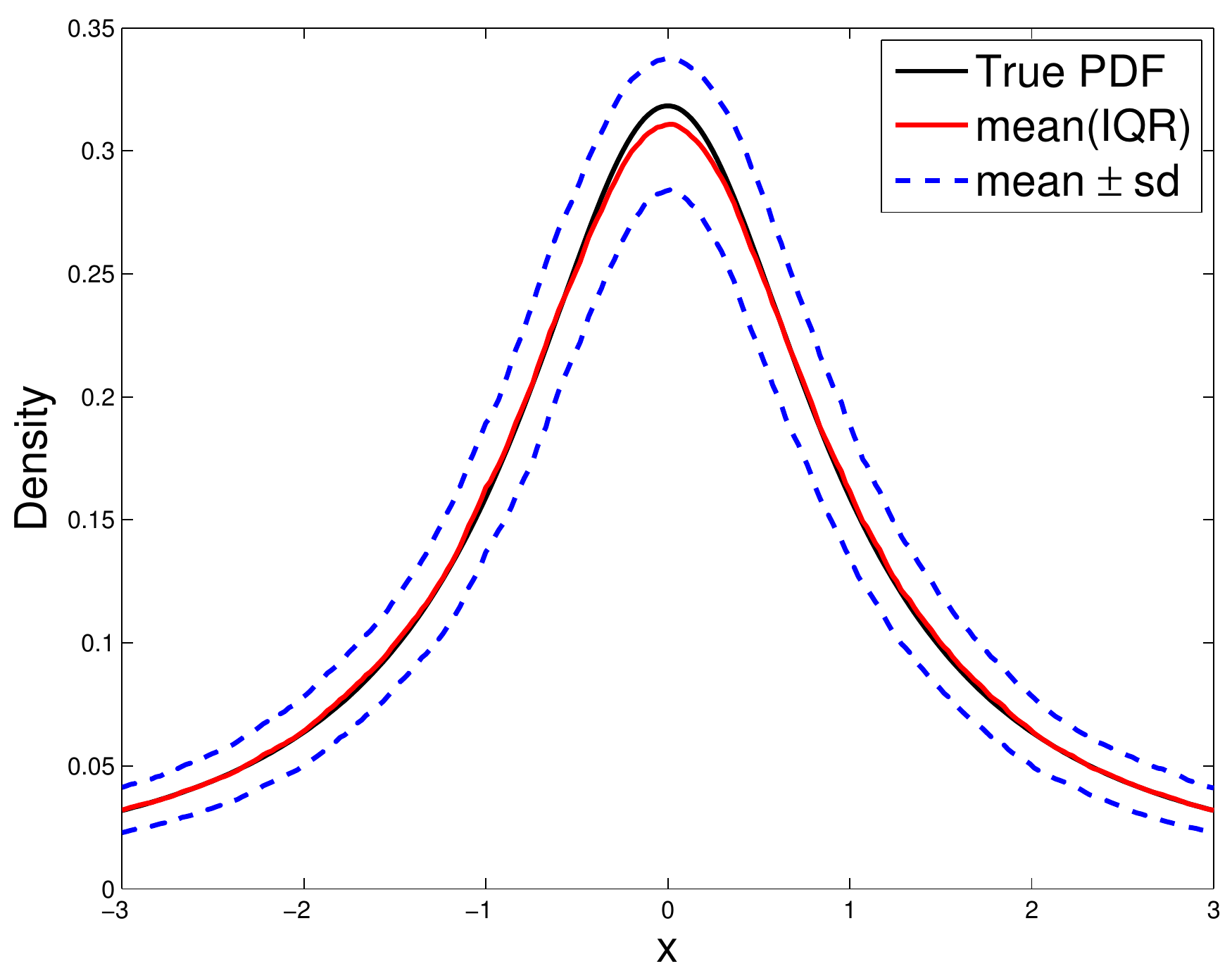}
\label{fig:cauchy_std_iqr}}
\caption{Pointwise mean and variance of the adaptive KDE and IQR based histogram methods.}
\label{fig:std_bias_others} 
\end{center}
\end{figure}

We now consider the pointwise bias and variance of 
MLD-DE. Given a fixed sample size, $N$, the bias and variance
are estimated by simulations over 1000 trials. Figure~\ref{fig:std_bias} shows the results; it shows 
pointwise estimates of the mean and the standard error of the density estimates plotted alongside the 
true densities. We see that the pointwise variance increases with the value
of the true density, while the bias is larger towards the corners of the estimation region. 
For comparison, Figure~\ref{fig:std_bias_others} shows analogous plots for the KDE and IQR 
histogram methods. 
In particular, for the beta density (Figure~\ref{fig:beta_std}), the
bias is smaller in the middle regions of the support of the
density. However, the bias is large near the boundary point $x=0$, where the
density has a discontinuity. Figure~\ref{fig:gmixture_std} shows the
corresponding results for the Gaussian mixture. Again, we see
a smaller variance in the tails of the density, but a larger bias
in the tails. As the variance increases with the density, we see larger
variances near the peaks than at the troughs.  The results improve
considerably with the optimal choice of $\alpha$
(Figure~\ref{fig:gmixture_std_opt_a}), with a significant decrease in
the bias.  Figure~\ref{fig:cauchy_std} shows the results for
the Cauchy density; these show a small bias in the tails but very low
variance.
\subsection{Effect of varying the tuning parameter $\alpha$}
The MLD-DE method depends on the parameter, $\alpha$, that controls the ratio of number of
subsets, $m_N$, to size, $s_N$, of each subset.  This is
similar to the dependence of histogram and KDE methods on a bandwidth
parameter. However, MLD-DE allows the use of different $\alpha$ 
at each point of estimation without affecting the
estimates at other points. This opens the possibility of flexible
adaptive density estimation. 

\begin{figure}[!h]
\begin{center}
\subfigure[Beta(1,4) distribution]
{
\includegraphics[width=0.35\textwidth]{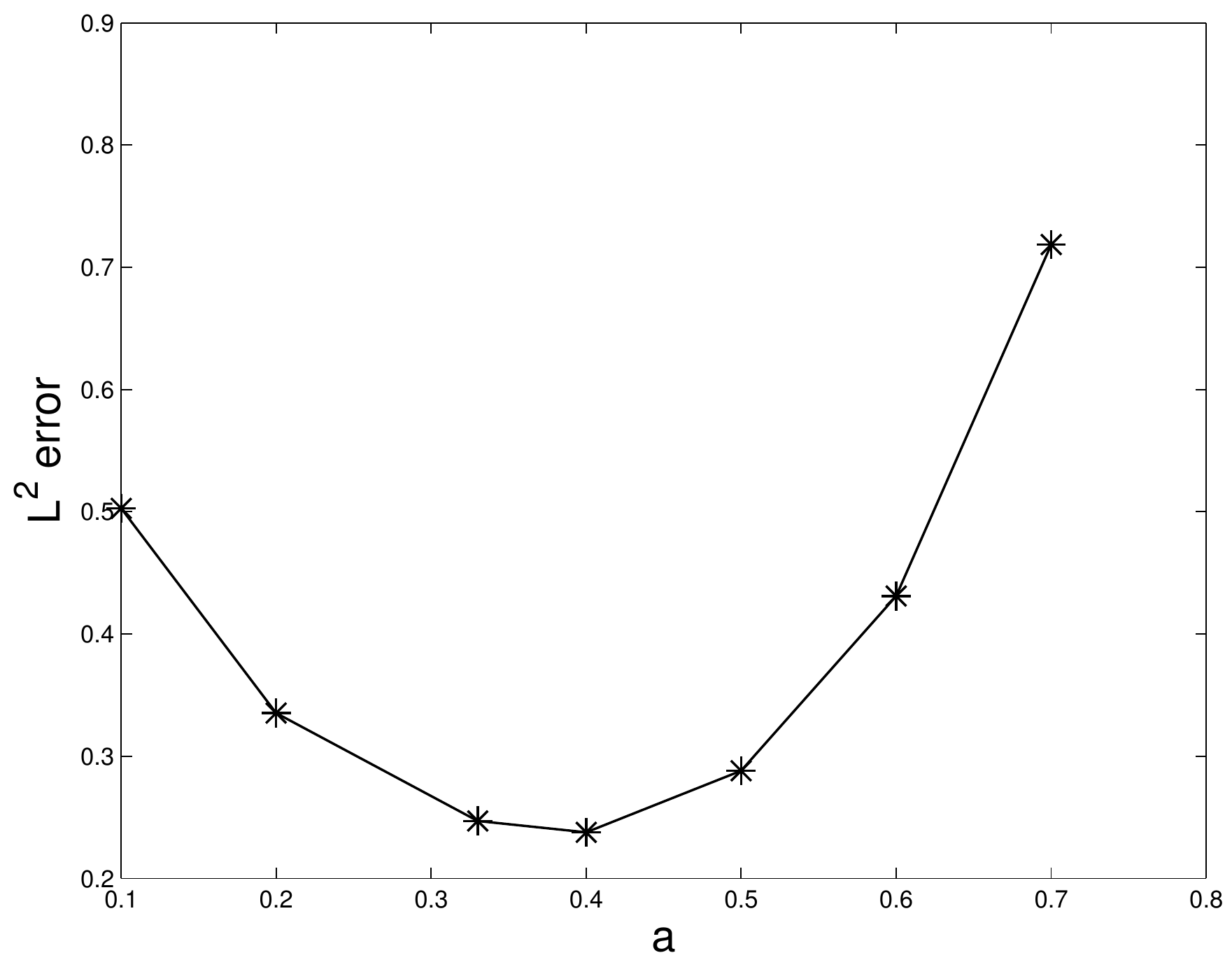}
\label{fig:beta_a}}
\subfigure[Gaussian Mixture]
{
\includegraphics[width=0.35\textwidth]{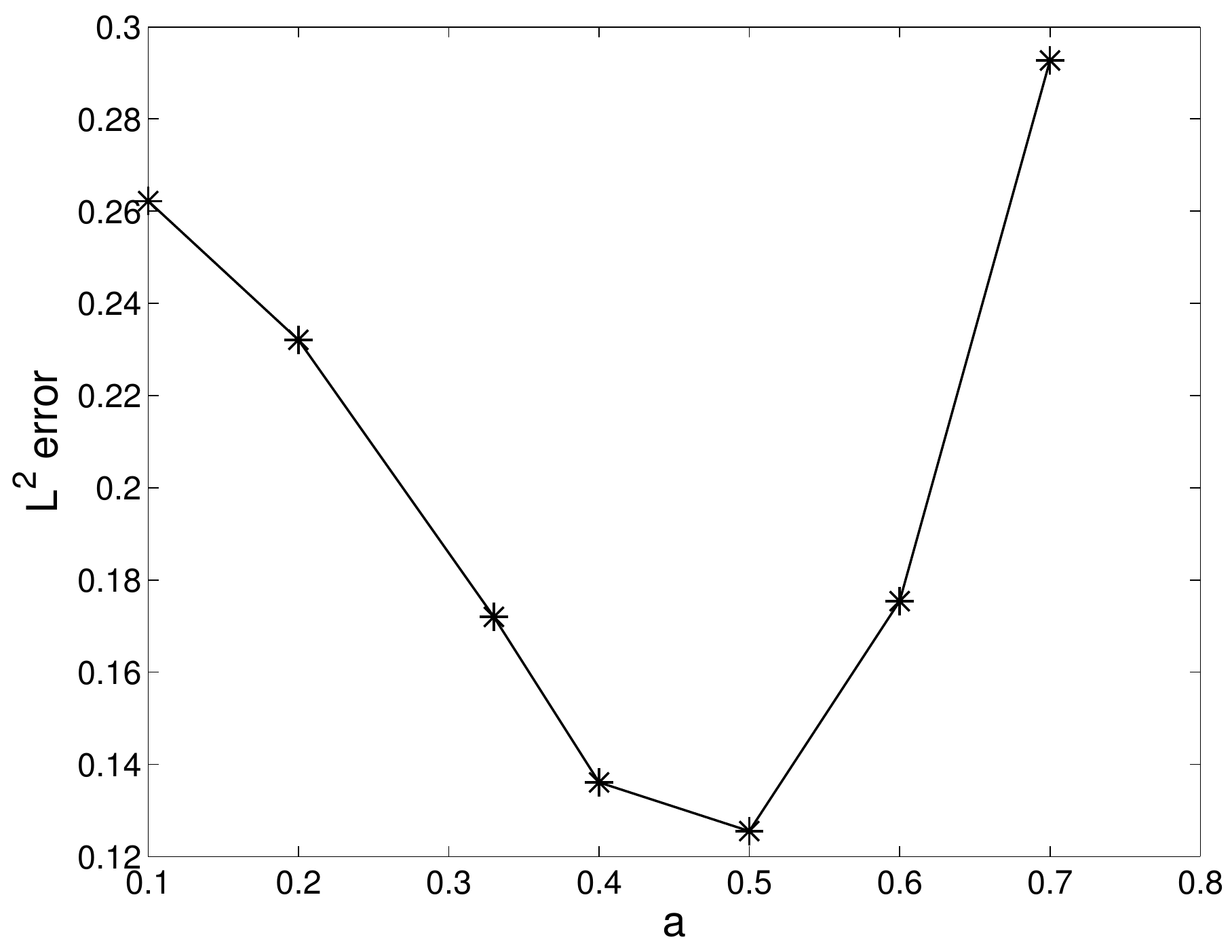}
\label{fig:gmixture_a}}
\subfigure[Cauchy(0,1) distribution]
{
\includegraphics[width=0.35\textwidth]{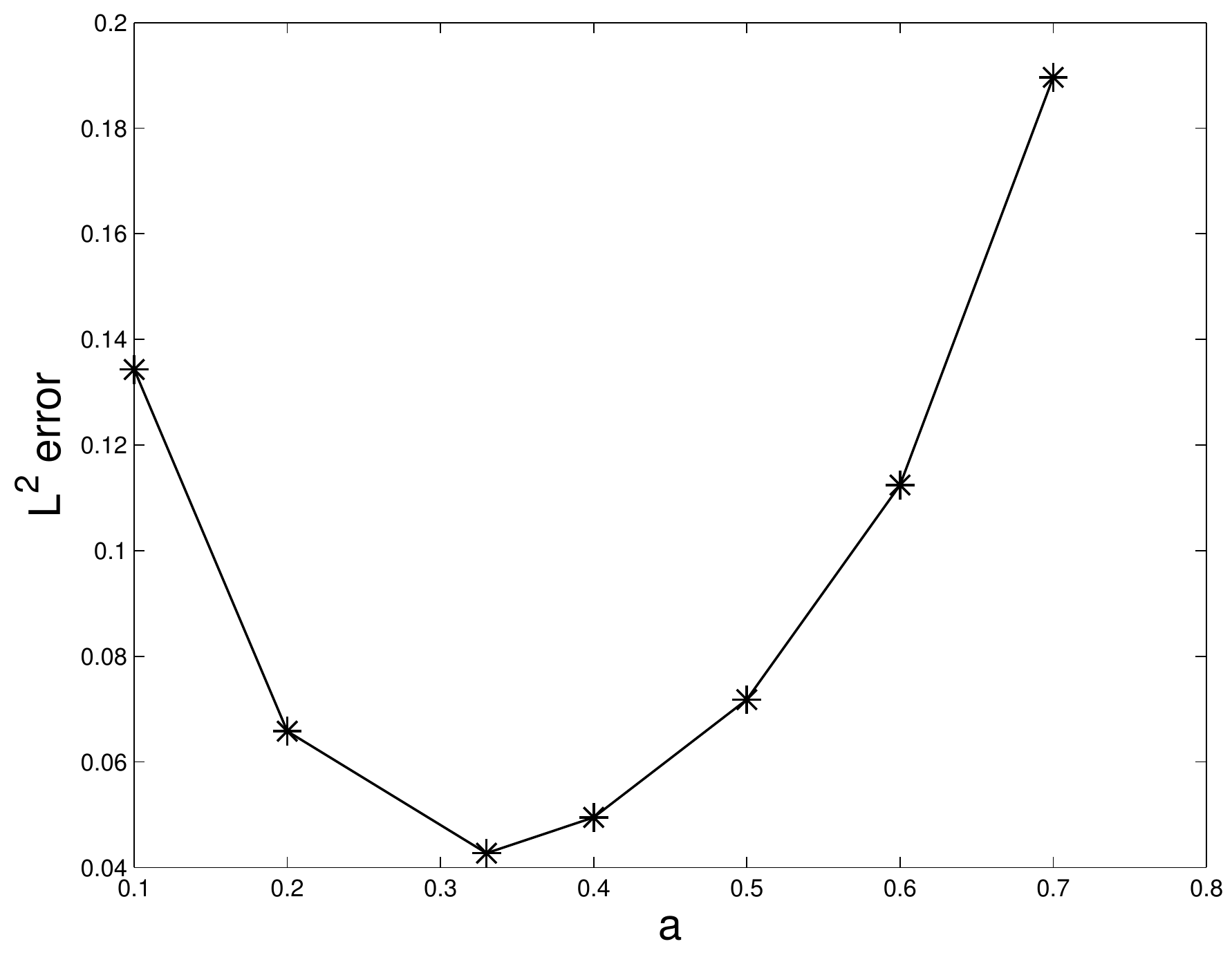}
\label{fig:cauchy_a}}
\caption{$L^{2}$-error versus the parameter $\alpha$ for various densities. The sample size
was fixed to $N = 1000$. A large $\alpha$ implies a small number of subsets $m_N$, but a 
large number of samples $s_N$ in each subset, while a smaller $\alpha$ implies the converse.}
\label{fig:error_vs_a} 
\end{center}
\end{figure}

To evaluate the effect of $\alpha$ on the 
$L^{2}$-error, simulations were performed using values of $\alpha$ that
increased from zero to one, with the total number of samples fixed to $N=1000$.
The simulations were done for the beta$(1,4)$, Gaussian
mixture and Cauchy$(0,1)$ distributions. Figure~\ref{fig:error_vs_a}
shows plots of the estimated $L^{2}$-error as a function of $\alpha$ for
the different densities. 
All the curves have a similar profile, with 
the error increasing sharply for $\alpha \geq 0.7$; so the plots only show
the errors for $\alpha \leq 0.8$. This indicates that, as we saw in Section \ref{sec:theory}, the
number of subsets must be larger than their size.
As we decrease $\alpha$ (i.e., increase the number of subsets), we see that
the error is less sensitive to changes in the parameter. Decreasing
$\alpha$ increases the bias, but keeps the variance low. In general, the
`optimal' value of $\alpha$ lies in between 0.2 and 0.6 for these
simulations, which further restricts the search range of any optimization
problem for $\alpha$.
\subsubsection*{An example of adaptive implementation}
An adaptive approach was used to improve MLD-DE estimates of the Cauchy distribution. The numerical results in Figure~\ref{fig:cauchy_std} indicate that 
there is a larger  bias in the tails of the distribution, while the theory indicates that the bias can be reduced by 
decreasing the number of subsets (correspondingly increasing the number of samples in each subset). The 
adaptive procedure used is as follows:
(1) A pilot density was first computed using MLD-DE with $\alpha = {1}/{3}$; (2) 
The points of estimation where the pilot density was within a fifth of the gap between the maximum and minimum 
density values from the minimum value (i.e., where the density was relatively small)
 were identified;
(3) The MLD-DE procedure was repeated with the value $\alpha={1}/{2}$ for those points of 
estimation.
\begin{figure}[!h]
\begin{center}
\subfigure[Mean Absolute Error]
{
\includegraphics[width=0.35\textwidth]{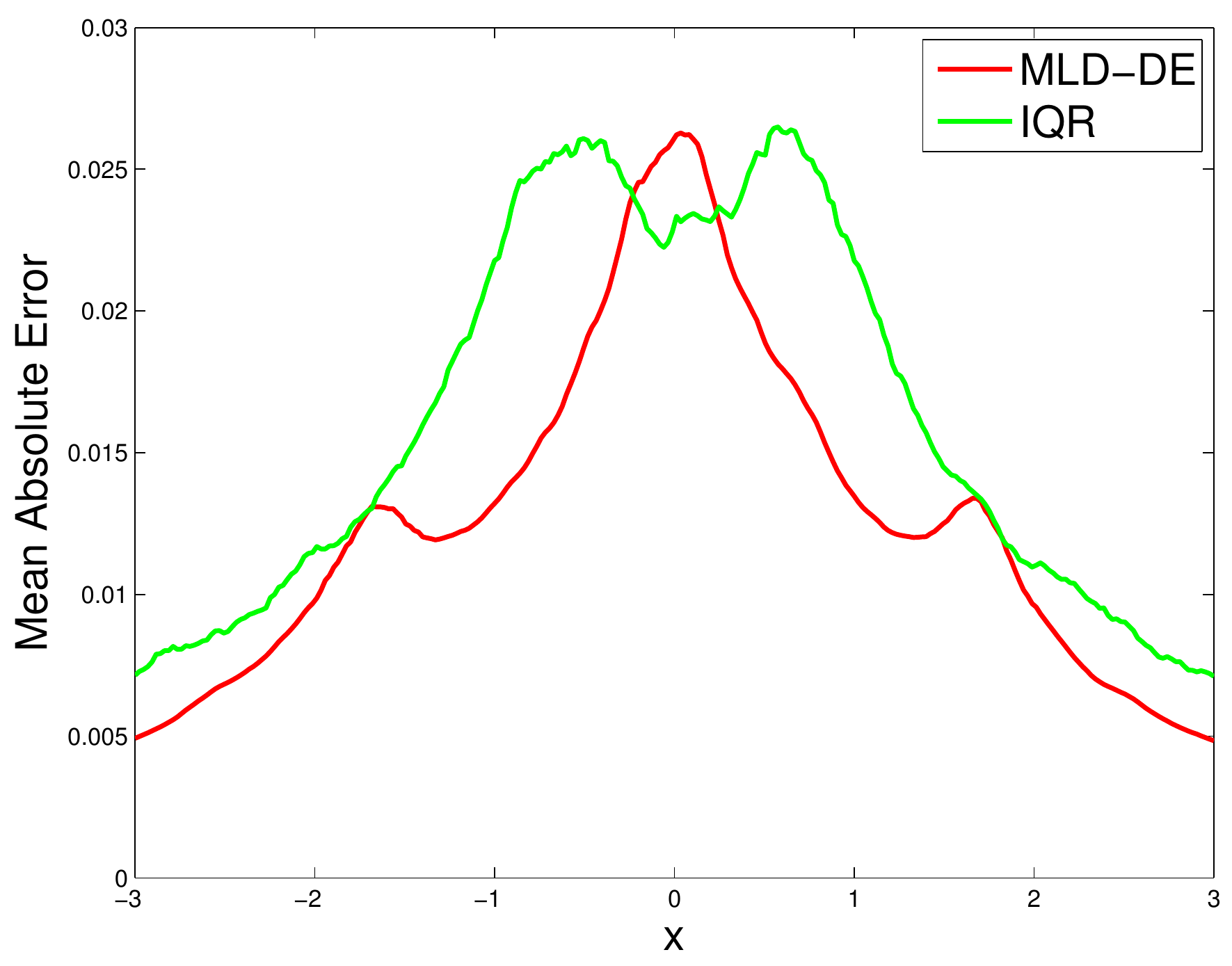}
\label{fig:mae_cauchy_adaptive}}
\subfigure[Bias and Variance Error]
{
\includegraphics[width=0.35\textwidth]{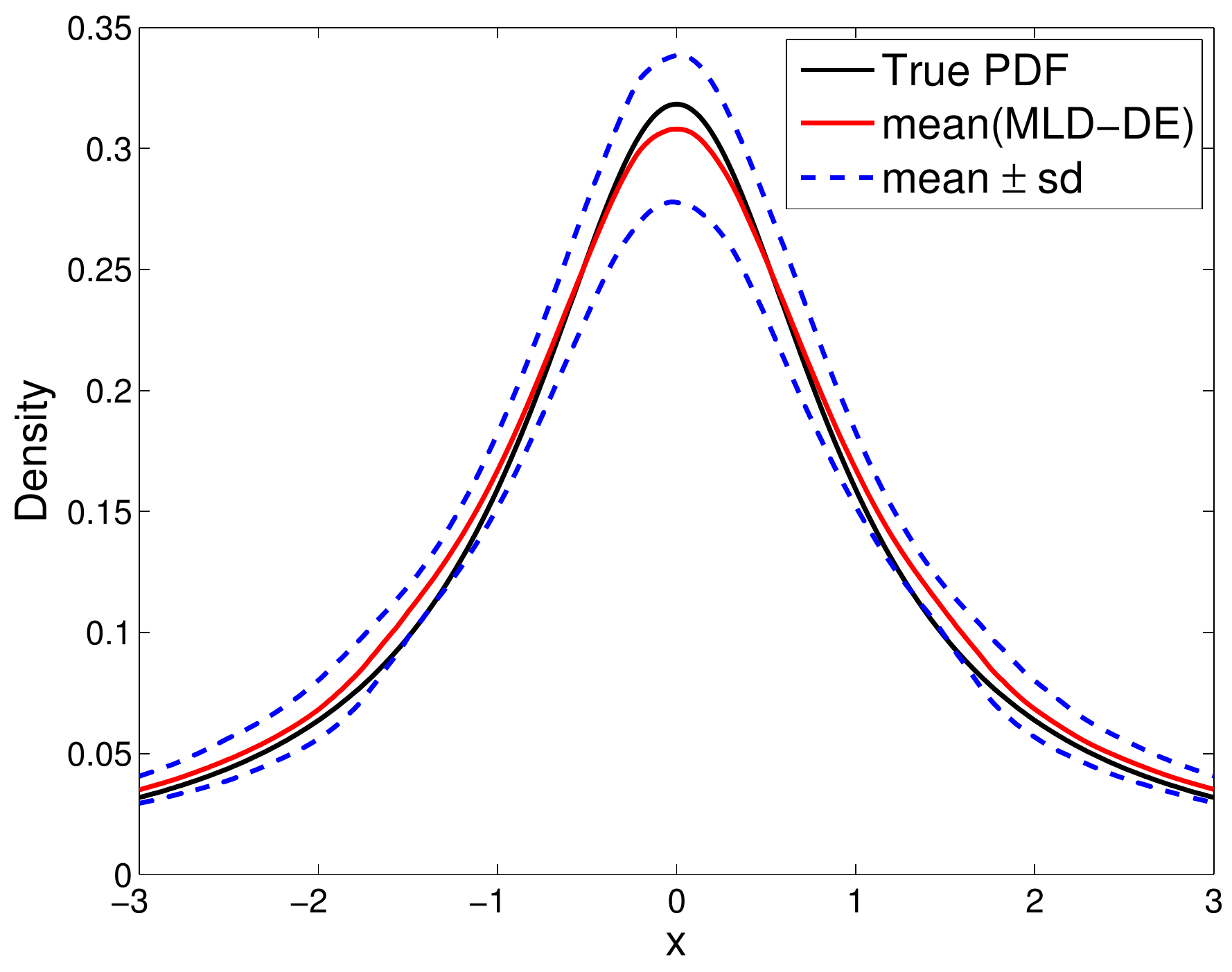}
\label{fig:std_cauchy_adaptive}}
\subfigure[Adaptive MLD Density]
{
\includegraphics[width=0.35\textwidth]{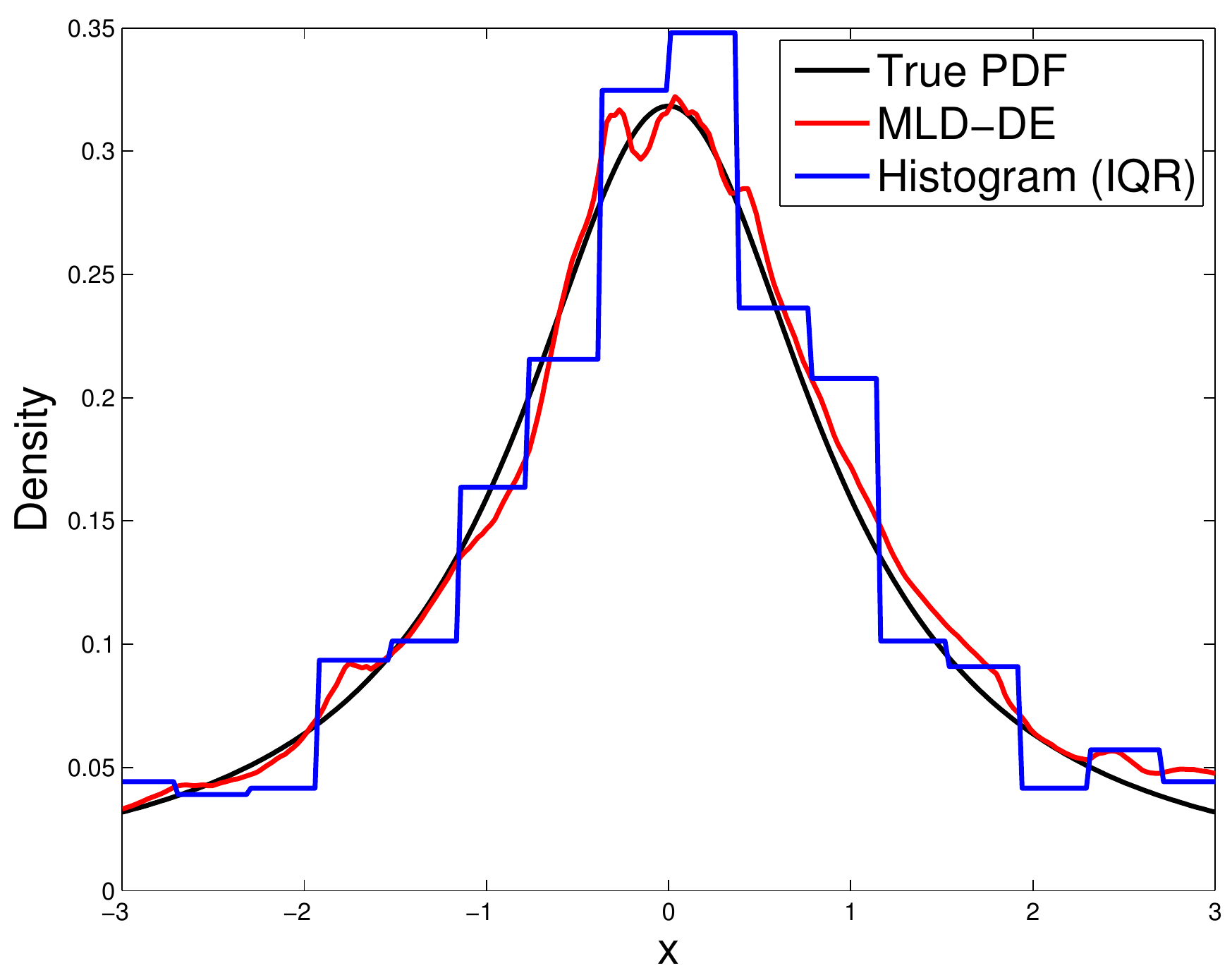}
\label{fig:comparison_cauchy_adaptive}}
\caption{Cauchy density estimation with the adaptive MLD-DE}
\label{fig:cauchy_adaptive} 
\end{center}
\end{figure}
Figure~\ref{fig:cauchy_adaptive} shows the results of this adaptive approach. We see that the 
bias 
has decreased significantly compared to that shown in the earlier plots for the non-adaptive 
approach. More sophisticated adaptive strategies can be employed with MLD-DE on account of 
its 
naturally adaptive nature, however a discussion of them is beyond the scope of this paper.

\section{Discussion and generalizations \label{sec:conc}}
We have presented a simple, robust and easily parallelizable method for one-dimensional density 
estimation. Like nearest-neighbor density estimators,
the method is based on nearest-neighbors but it offers the advantage of providing smoother density estimates,
and has parallel complexity $\orderof{N^{{1}/{3}}}$ .
Its tuning parameter is the number of subsets in which the original sample is divided.
Theoretical results concerning the asymptotic distribution of the estimator were developed and 
its MSE was analyzed to determine a globally optimal split of the original sample 
into subsets. Numerical experiments illustrate that the method can recover different types of densities, 
including the Cauchy density, without the need for special kernels or bandwidth selections. 
Based on a heuristic analysis of high bias in low-density regions, an adaptive implementation that
reduces the bias was also presented. Further work 
will be focused on more sophisticated adaptive schemes for one-dimensional density estimation and 
extensions to higher dimensions. We present here a brief overview of a higher 
dimensional extension of MLD-DE. Its generalization is straightforward but its convergence is
usually not better than that of histogram methods. To see why, we consider the bivariate case.
Let $(X,Y)$ be a random vector with PDF $f(x,y)$, and let $h(x,y)$ and $H(x,y)$ be the PDF
and CDF of $(|X|,|Y|)$. It is easy to see that
\(
h(0,0) = 4 f(0,0).
\)
In addition, let $q(t)=H(t,t)$, then
\(
q'(t) = \int_0^t h(t,y)\,dy + \int_0^t h(x,t)\,dx.
\)
It follows that (assuming continuity at $(0,0)$),
\(
q''(0)=\lim_{t\to 0} q'(t)/t = 2\, h(0,0).
\)
Let $\Xb_1=(X_1,Y_1),\ldots,\Xb_N=(X_N,Y_N)$ be iid vectors and define $U_i$ to be the product
norm of $\Xb_i$:
\(
U_i = \|\Xb_i\|_\otimes = \max\{\,|X_i|,\,|Y_i|\,\},
\)
and $Z = U_{(1)}$. Then
\begin{eqnarray*}
\Prob(\,Z>t\,) &=& \Prob( \,\|\Xb_1\|_\otimes >t,\ldots, \|\Xb_N\|_\otimes >t\,)=\Prob( \,\|\Xb_1\|_\otimes >t\,)^N\\
&=& [\,1-\Prob(\,\|\Xb_1\|_\otimes \leq t\,)\,]^N = 
[\,1-\Prob(\,|X_1| \leq t,|Y_1| \leq t\,)\,]^N\\
&=& [\,1-q(\,t\,)\,]^N.
\end{eqnarray*}
Let $Q$ be the inverse of the function $q$. It is easy to check that
\(
Q'(z) = 1/q'(Q(z)).
\)
Proceeding as in the 1D case, we have
\begin{eqnarray*}
\Ex (\,Z^2\,) &=& 2\!\!\int_0^\infty \hspace{-.3cm}t\,\Prob(\,Z>t\,)\,dt 
= 2\!\!\int_0^\infty\hspace{-.3cm} t\,[\,1-q(\,t\,)\,]^N\,dt
= \int_0^1 (Q^2(z))' \,(1-z)^N\,dz.
\end{eqnarray*}
Therefore
\(
\Ex [\,(N+1)\,Z^2\,] =  \int_0^1 (Q^2(z))' \,\delta_N(z)\,dz,
\)
and by the results in Section \ref{sec:theory},
\[
\lim_n \Ex[\,(N+1)\,Z^2\,] = (Q^2(z))'|_{z=0}={1}/{h(0,0)} ={1}/{(4\,f(0,0))}. 
\]
Furthermore,
\[
\Ex [\,(N+1)\,Z^2\,] =  
\frac{1}{h(0,0)} + \frac{1}{N+2}\int_0^1 (Q^2(z))'' \,\delta_{N+1}(z)\,dz.
\]
But, unlike in the 1D case, this time we have
\(
\lim_{z\to 0}\, (Q^2(z))'' = {q^{(4)}(0)}/{(3 q''(0))}\neq 0,
\)
and this makes the convergence rates closer to those of histogram methods.

\smallskip
\noindent{\bf Acknowledgments.}
We thank Y. Marzouk for reviewing our proofs, and 
D. Allaire, L. Ng, C. Lieberman and R. Stogner for helpful discussions. 
The first and third authors acknowledge the support of the the DOE Applied Mathematics Program, 
Awards DE-FG02-08ER2585 and DE-SC0009297, as part of the DiaMonD Multifaceted Mathematics Integrated Capability Center.

\bibliographystyle{apalike}
\bibliography{paper}

\appendix
\section{Proofs}\label{sec:proofs}
\noindent
\textbf{Proof of Lemma \ref{lemma:gprop}:}
(i) Since $f$ has right and left limits, $f(x_*^+)$ and $f(x_*^-)$, at $x_*$, we may
re-define $f(x_*) = (f(x_*^+) +f(x_*^-))/2$.  It then follows that
\[
\lim_{y\to 0^+}g(y) = f(x_*^+) +f(x_*^-)= 2f(x_*) = g(0),
\]
and therefore $g$ is right-continuous at zero.
(ii) If $f$ has right and left derivatives, $f'(x_*^+)$ and $f'(x_*^-)$, at $x_*$, then
\(
\lim_{y\to 0^+}({g(y)-g(0)})/{y}= f'(x_*^+)-f'(x_*^-)
\)
and therefore $g'(0)$ exits, and $g'(0)=0$ if $f$ is differentiable at $x_*$.
$\qedsymbol$

\medskip
The proof of Proposition \ref{prop:convlim} makes use of the elementary fact that the functions
$\delta_N(z) = (N+1)(1-z)^N$, $z\in [0,1]$, define a sequence of Dirac functions. That is,
for every $N\in \mathbb{N}$: (i) $\delta_N\geq 0$;
(ii) $\int_0^1\delta_N(z)\,dz = 1$; and (iii) for any $\varepsilon>0$ and 
$\delta\in (0,1) $, there
is an integer $N_0$ such that
\(
\int_{\delta}^1\delta_N(z)\,dz <\varepsilon
\)
for any $N\geq N_0$.
\medskip

\noindent
\textbf{Proof of Lemma \ref{lemma:ordmom}:}
The results follow from straightforward applications of the tail formula for the moments of a non-negative
random variable. For (i) we have
\[
\Ex \Xone = \int_0^\infty \Prob (\Xone > t)\,dt = \int_0^\infty (1-G(t))^N\,dt.
\]
Using the change of variable $z=G(t)$ leads to
\(
\Ex \Xone = \int_0^1 Q'(z)\,(1-z)^N\,dz,
\)
and \eqref{eq:EXone} follows. Version \eqref{eq:EXonev1} follows from \eqref{eq:EXone} using 
integration by parts, while version \eqref{eq:EXonev1} follows using two integration
by parts and the fact that $g'(0)=0$. For (ii) we have something similar,
\begin{eqnarray*}
\Ex[\Xone^2] &=& \int_0^\infty2 t\,\Prob (\Xone > t)\,dt = \int_0^\infty 2t\, (1-G(t))^N\,dt\\
& = & \int_0^1 2 \,Q(z) Q'(z) \,(1-z)^N\,dz,
\end{eqnarray*}
and therefore
\begin{eqnarray*}
(N+1)^2\,\Ex[\Xone^2] &=&  (N+1)\int_0^1 (Q^2(z))' \,\delta_N(z)\,dz\\
&=&  \left(\frac{N+1}{N+2}\right)\int_0^1 (Q^2(z))''\,\delta_{N+1}(z)\,dz.
\end{eqnarray*}
The last equation follows from integration by parts. $\square$
\medskip

\noindent
\textbf{Proof of Proposition \ref{prop:convlim}:}
Assume first that $H(0)\neq 0$. Let $\varepsilon>0$. By continuity at 0, there is $\eta\in (0,1)$ such that 
$|H(z)-H(0)|< \varepsilon/3$ if $0\leq z<\eta$. Also, by the properties of $\delta_N$ and, 
because $(N+1)/(N-m+1)\to 1$ as $N\to \infty$, there is an integer $N_0$ such that for $N>N_0$,
\[
\int_{\eta}^1\!\!\delta_N(z)\,dz < \min\{{\varepsilon}/{(3 |H(0)|)},{\varepsilon}/{(6C)}\}
\,\,\mbox{and}\,\, (N+1)/(N-m+1)\leq 2.
\]
Then,
\begin{eqnarray*}
\left|\int_0^1 \!\!H(z) \,\delta_N(z)\,dz- H(0)\right| &\leq & 
\int_0^{\eta}\hspace{-.2cm} |H(z)-H(0)|\,\delta_N(z)\,dz
+ \int_{\eta}^1\hspace{-.2cm}|H(z)|\,\delta_N(z)\,dz \\
& & +  \quad |H(0)|\int_{\eta}^1\hspace{-.2cm} \delta_N(z)\,dz\\
&\leq & \varepsilon/3 + \varepsilon/3 + \int_{|z|\geq \eta} \!\!\!\!|H(z)|\,\delta_N(z)\,dz\\
&\leq & \varepsilon/3 + \varepsilon/3 + C\!\left(\frac{N+1}{N-m+1}\right)\!\!
\int_{\eta}^1 \hspace{-.2cm}\delta_{N-m}(z)\,dz\leq \varepsilon
\end{eqnarray*}
for $N>N_0+m$. The proof for $H(0)=0$ is analogous. $\square$
\medskip

\noindent
\textbf{Proof of Proposition \ref{prop:mselim}:}
(i) Since for a fixed $s_N$, $X^{(1)}_{(1),s_N},\ldots,X^{(m_N)}_{(1),s_N}$ is an iid sequence, it follows from Corollary
\ref{cor:momlims}(i) that
\[
\Ex\,\finvh_{N} =  ({1}/{m_N})\sum_{k=1}^{m_N}\Ex[\,(s_N+1)X^{(k)}_{(1),s_N}\,]
= \Ex[\,(s_N+1)X^{(1)}_{(1),s_N}\,]
\to  {1}/{g(0)} 
\]
as $N\to \infty$. For the second moment we have (for simplicity we define $a_N=s_N+1$),
\begin{eqnarray*}
\Ex[\,\finvh_{N}^2\,]& =&  \frac{1}{m_N^2}\sum_{k=1}^{m_N}\Ex[\,(a_N X^{(k)}_{(1),s_N})^2\,]
+\frac{1}{m_N^2}\sum_{j\neq k}\Ex(\,a_N X^{(j)}_{(1),s_N}\, )\Ex(\,a_N X^{(k)}_{(1),s_N}\,  )\\
&=&\frac{1}{m_N}\Ex[\,(a_N X^{(1)}_{(1),s_N})^2\,]+\left(\,\frac{m_N-1}{m_N}\,\right)
[\,\Ex(\,a_N X^{(1)}_{(1),s_N}\,  )\,]^2.
\end{eqnarray*}
The variance and MSE are thus given by
\begin{eqnarray*}
\Var[\,\finvh_{N}^2\,]& =&  ({1}/{m_N})\Ex[\,(a_N X^{(1)}_{(1),s_N})^2\,]-
({1}/{m_N})[\,\Ex(\,a_N X^{(1)}_{(1),s_N}\,  )\,]^2\\
\mse(\,\finvh_{N}\,) &=& \frac{1}{m_N}\Ex[\,(a_N X^{(1)}_{(1),s_N})^2\,] + \left(\,\frac{m_N-1}{m_N}\,\right)
[\,\Ex(\,a_N X^{(1)}_{(1),s_N}\,  )\,]^2\\
& & -\frac{2}{g(0)}\Ex(\,a_N X^{(1)}_{(1),s_N}\,  ) + \frac{1}{g(0)^2}.
\end{eqnarray*}
By \eqref{eq:limfirst} and \eqref{eq:limsec}, the MSE converges to zero as $N\to\infty$,
Hence, \eqref{eq:mselim} follows. \\
(ii) Since limit \eqref{eq:fidistlim} implies \eqref{eq:fdistlim} by Cramer's $\delta$-method, it is sufficient
to prove \eqref{eq:fidistlim}.
Define $\mu_N = \Ex X_{(1),s_N}^{(k)}$,\, $Z_{k,s_N}= \,X_{(1),s_N}^{(k)}-\mu_N$, and $Y_{k,N}=a_N\,Z_{k,s_N}/\sqrt{m_N}$ for
$k\leq m_N$, with  $Y_{k,N}=0$, otherwise.
The variables $Y_{1,N},\ldots,Y_{N,N}$ are independent, zero-mean and, by Corollary \ref{cor:momlims}(iii),
\begin{equation}\label{eq:LFC1}
\sum_{k=1}^N \Ex (\,Y_{k,N}^2\,) = \Var(\,a_N X_{(1),s_N}^{(1)}\,)\to 1/g(0)^2,
\end{equation}
as $N\to\infty$. Fix $\varepsilon>0$. We show that the following Lindeberg condition is satisfied:
\begin{equation}\label{eq:LFC2}
\sum_{k=1}^N \Ex (\,Y_{k,N}^2 \,I_{Y_{k,N}^2>\varepsilon^2} \,) = 
\Ex (\,a_N^2Z_{1,N}^2 \,I_{a_N^2 Z_{1,N}^2>\varepsilon^2{m_N}} \,)\to 0.
\end{equation}
To see this, note that since $X_i\geq 0$, we have
\(
a_N^2 Z_{1,s_N}^2 \leq a_N^2X_{(1),s_N}^2 + a_N^2\mu_N^2. 
\)
Since $a_N\mu_N$ has a finite limit, the difference $\varepsilon m_N - a_N^2\mu_N^2$
is positive for $N$ larger than some integer $N_1$. For simplicity, define 
$c_N^2 =\varepsilon^2 m_N - a_N^2\mu_N^2$.
We then have
\[
\Ex (\,a_N^2Z_{1,N}^2 \,I_{a_N^2Z_{1,N}^2>\varepsilon^2 m_N} \,)\leq
\Ex(\,a_N^2X_{(1),s_N}^2I_{a_N^2 X_{(1),s_N}^2>c_N^2}\,) + 
a_N^2\mu_N^2\Prob(a_N^2X_{(1),s_N}^2 >c_N^2).
\]
Since $m_N/s_N\to \infty$, it follows that $c_N/a_N\to \infty$ and therefore
$\Prob(\, X>c_N/a_N\,)<1/2$ for $N$ larger than an integer $N_2$. Hence, for $N>\max\{N_1,N_2\}$,
\begin{eqnarray*}
a_N^2\mu_N^2\Prob(a_N^2X_{(1),s_N}^2 >c_N^2) \!\!\!&=&\!\! a_N^2 \mu_N^2\Prob(\, X>c_N/a_N\,)^{s_N}
\leq a_N^2 \mu_N^2/2^{s_N}\\
\Ex(\,a_N^2X_{(1),s_N}^2I_{a_N^2 X_{(1),s_N}^2>c_N^2}\,) \!\!\!&= & 
2\!\!\int_0^\infty\hspace{-.3cm} t\,\Prob(a_N^2X_{(1),s_N}^2I_{a_N^2 X_{(1),s_N}^2>c_N^2}>t)\,dt\\
\!\!\!&\leq & 2\!\!\int_0^\infty\hspace{-.3cm} t\,\Prob(\,a_N X_{(1),s_N}>t,a_N X_{(1),s_N}>c_N\,)\,dt\\
\!\!\!&=& 2\Prob(\, X_{(1),s_N}>c_N/a_N\,)\!\int_0^{c_N} \hspace{-.3cm}t\,dt +
2\!\!\int_{c_N}^\infty\hspace{-.3cm} t \,\Prob(\, X_{(1),s_N}>t/a_N\,)\,dt.
\end{eqnarray*}
The tail condition (with $C>0$ and integer $k>0$) on the 
last integral yields
\begin{eqnarray*}
\Ex(\,a_N^2X_{(1),s_N}^2I_{a_N^2 X_{(1),s_N}^2>c_N^2}\,)
&\leq & c_N^2\Prob(\, X_{(1),s_N}>c_N/a_N\,)+C a_N^2\!\!\int_{G(c_N/a_N)}^1\hspace{-.8cm} (1-z)^{s_N-k}\,dz\\
&=& c_N^2\Prob(\, X>c_N/a_N\,)^{s_N}+\frac{C a_N^2}{a_N+k}\Prob(\, X>c_N/a_N\,)^{a_N-k}\\
&\leq &\frac{c_N^2}{2^{s_N}}+\frac{C\,a_N^2}{(a_N+k)\,2^{a_N-k}}.
\end{eqnarray*}
Since the right hand-side converges to zero, \eqref{eq:LFC2} follows. By Lindeberg-Feller's theorem,
\eqref{eq:LFC1} and \eqref{eq:LFC2} imply that
\begin{equation}\label{eq:distlimmu}
\sqrt{m_N}\,[\,\finvh_{N} - \Ex\finvh_{N}\,] \arrL N(0,1/g(0)^2).
\end{equation}
On the other hand, by \eqref{eq:limfirstv2},
\begin{equation}\label{eq:mulimit}
\sqrt{m_N}\,[\,a_N \mu_N - 1/g(0)\,]\to 0
\end{equation}
because $\sqrt{m_N}/s_N^2\to 0$. Combining \eqref{eq:limfirstv2} and \eqref{eq:mulimit} 
yields \eqref{eq:fidistlim}. Note that since $s_N^2 > \sqrt{m_N}$, it follows that the ME of 
$\finvh_N$ is $\orderof{1/m_N}$, and using a simple Taylor expansion one also finds that the
MSE of $\fh_N$ is also $\orderof{1/m_N}$.
$\square$
\end{document}